\documentclass[pre,reprint,showpacs,amsmath,amssymb]{revtex4-1} 
\usepackage{bm}
\usepackage{graphicx}
\usepackage{amsmath}
\usepackage{eqnarray}
\usepackage{latexsym}
\usepackage{amsfonts}
\usepackage{psfrag}
\usepackage{float}
\usepackage{dcolumn}
\usepackage{epsfig}
\usepackage{graphics}
\usepackage{amssymb}
\usepackage{textcomp}
\usepackage{subfigure}

\usepackage{color}
\usepackage[dvipsnames]{xcolor}

\include{macros}

\newcommand{\subCS}{{_\mathrm{CS}}}
\newcommand{\subG}{{_\mathrm{G}}}

\newcommand{\bigO}{\mathcal{O}}
\newcommand{\delT}{\delTw}
\newcommand{\dif}{\mathcal{D}}

\newcommand{\difp}{\dif_\mathrm{p}}
\newcommand{\difs}{\dif_\mathrm{s}}
\newcommand{\difw}{\dif_\mathrm{w}}
\newcommand{\dwater}{h}
\newcommand{\Plaser}{\mathcal{P}}
\newcommand{\Pth}{P_\mathrm{th}}
\newcommand{\Rc}{R_\mathrm{c}}
\newcommand{\sig}{\sigma}
\newcommand{\Sc}{S_\mathrm{c}}
\newcommand{\SM}{S.M.~\cite{sm}}
\newcommand{\To}{T_0}
\newcommand{\taul}{\hat{\tau}}
\newcommand{\taulcs}{\taul_\subCS }
\newcommand{\taulbare}{\taul_\subG}
\newcommand{\taulinf}{\taul_\infty}
\newcommand{\taulo}{\taul_o}

\newcommand{\taus}{\bar{\tau}}
\newcommand{\tauscs}{\taus_\subCS }
\newcommand{\tausbare}{\taus_\subG}
\newcommand{\tauee}{{\tau_\mathrm{e-e}}}

\newcommand{\theating}{\tau}

\newcommand{\tpulse}{t_\mathrm{pul}}
\newcommand{\Vc}{V_\mathrm{c}}

\newcommand{\sbr}[2]{{#1_{\mathrm{#2}}}}
\newcommand{\ce}{\sbr{c}{e}}
\newcommand{\cp}{\sbr{c}{p}}
\newcommand{\cs}{\sbr{c}{s}}
\newcommand{\cw}{\sbr{c}{w}}
\newcommand{\delTein}{\sbr{\Delta T}{e,in}}
\newcommand{\delTw}{\sbr{\Delta T}{w}}

\newcommand{\ks}{\sbr{k}{s}}
\newcommand{\kw}{\sbr{k}{w}}

\newcommand{\siges}{\sbr{\sig}{es}}
\newcommand{\sigps}{\sbr{\sig}{ps}}
\newcommand{\sigsw}{\sbr{\sig}{sw}}
\newcommand{\sigpw}{\sbr{\sig}{pw}}
\newcommand{\kappacs}{\kappa_\subCS }
\newcommand{\kappabare}{\kappa_\subG }

\newcommand{\Te}{\sbr{T}{e}}
\newcommand{\Tp}{\sbr{T}{p}}
\newcommand{\Ts}{\sbr{T}{s}}

\newcommand{\Tw}{\sbr{T}{w}}

\newcommand{\Vs}{\sbr{V}{s}}

\newcommand{\modif}[1]{#1}

\usepackage{etoolbox}
\usepackage{xstring}
\newcommand\IfStringInList[2]{\IfSubStr{,#2,}{,#1,}}

\begin{document}

\title{Enhanced heat transfer with core-shell metal dielectric nanoparticles}

\date{\today}

\author{Ali Alkurdi}
\author{Julien Lombard}
\altaffiliation{also at: Departamento de Fisica y Quimica Teorica, Faculdad de Quimica, Universidad Autonomà de Mexico, Mexico, Mexico}
\altaffiliation{*These authors contribute equally to this work}
\author{Fran\c{c}ois Detcheverry}
\author{Samy Merabia}
\affiliation{Univ. Lyon, Universit{\'e} Claude Bernard Lyon 1, CNRS, Institut Lumi{\`e}re Mati{\`e}re, F-69622, Villeurbanne, France}

\begin{abstract}
Heat transfer from irradiated metallic nanoparticles is relevant to a broad array of applications ranging from water desalination to photoacoustics. 
The efficacy of such processes relies on the ability of these  nanoparticles 
to absorb the pulsed illuminating light and to quickly transfer energy to the environment.  
Here we show that compared to homogeneous gold nanoparticles having the same size, 
gold-silica core-shell nanoparticles enable heat transfers to liquid water that are faster.  
We reach this conclusion by considering both analytical and numerical calculations. 
The key factor explaining enhanced heat transfer 
is the direct interfacial coupling between metal electrons and silica phonons. 
We discuss how to obtain fast heating of water in the vicinity of the particle 
and show that optimal conditions involve nanoparticles with thin silica shells irradiated by ultrafast laser pulses. 
Our findings should serve as guides for the optimization of thermoplasmonic applications of core-shell nanoparticles.      
\end{abstract}


\maketitle

\section{Introduction}

Metallic nanoparticles have a remarkable ability 
to  absorb an electromagnetic radiation in the vicinity of the plasmon resonance wavelength, 
which lies in the visible or near infra red domains for common nanostructures~\cite{Bohren1983}. 
This phenomenon originates in the collective electromagnetic response of the metal conduction electrons which yields local field enhancement. 
In turn, this localized exaltation may be exploited 
to realize heat sources at the nanoscale~\cite{Govorov2007,Baffou2011,Baffou2012}. 
By tailoring the nanoparticle size and composition, 
the resonance wavelength may be tuned, which is advantageous if electromagnetic absorption or scattering from the nanoparticle environment has to be minimized. 

This sensitivity combined with the biocompatibility of gold nanoparticles (GNPs) have been exploited in different contexts, including thermally induced photochemistry~\cite{Burda2005}, phase transition~\cite{Richardson2006}, material growth~\cite{Cao2007}, and nanomedicine~\cite{Dreaden2012,Qin2012}. 
The latter involves  thermal imaging~\cite{Chen2011}, diagnosis, photoacoustics~\cite{Prost2015}, drug release~\cite{Han2007,McNeil2011}, 
and cancer therapy~\cite{Day2009,Alexis2010}, 
in which GNPs have been used to generate hyperthermia inside tumor cells leading to their destruction. 
In these applications, the efficiency of the treatment relies on the absorption contrast between the cancerous and the healthy tissues~\cite{Baffou2012}, which is realized by working in the so-called transparency window between $700$-$900\,$nm 
wavelength where the absorption of human tissues is minimum~\cite{Lal2008}.

Alongside GNPs, metal-dielectric core-shell nanoparticles (CSNP) have also been investigated as nano-source of heat with applications in cancer therapy~\cite{Hirsch2003}, optical enhancement~\cite{Meng2017,xu2005}, and solar energy assisted water purification~\cite{Neumann2012}, bringing in some cases clear benefits. 
One such instance is photoacoustics where the rapid heating of illuminated particles 
triggers an acoustic or stress wave induced by thermal expansion~\cite{Prost2015}.  
The efficiency of this process is intimately related to the ability of the nanoparticles to transfer heat in their environment. 
Recently, Chen et {\em al.}~\cite{Chen2011,Chen2012} used gold nanorods coated with silica to assess the ability of these nano-objects to enhance the contrast in photoacoustic imaging. 
They found that compared to bare gold nanorods, 
silica-coated  nanorods produce  photoacoustic signals which are  about three times higher. 
This enhancement is shown not to result from changes in the absorption cross section due to the silica coating. 
Rather, it is ascribed to a change in interface thermal conductivity induced by the silica shell. 
This interpretation is consistent with earlier experimental investigations of Hu et~{\em al.}~\cite{Hu2003}, which reported faster heat dissipation for gold nanoparticles coated with a silica shell.
The underlying mechanism, however, has not been elucidated so far.
 

In a different context, several experimental studies investigated heat transport at metal/dielectric interfaces, at picosecond time scales. 
Recent transient thermoreflectance experiments with a gold film on silicon and silica substrate~\cite{Hopkins2009,guo2012} 
concluded on the existence of a direct electron-substrate energy transfer. Indeed, there are at least two mechanisms according to which electrons may heat up the substrate: non-equilibrium electron-phonon exchange in the metal which then transfers energy to the substrate through phonon-phonon coupling~\cite{majumdar2004,jones2013,hopkins2013,wang2012b,singh2013}; 
and direct electron-phonon coupling through the interface~\cite{huberman1994,sergeev1998,mahan2009,zhangh2013}.  
\modif{While the effect of the electron-phonon coupling in the bulk metal is to increase the thermal boundary conductance~\cite{majumdar2004,lu2018}, 
the direct electron-phonon coupling provides another channel for heat to flow across the metal-dielectric interface~\cite{Lombard2015}.}

\modif{
This energy transfer through interfacial electron/phonon conductance 
was predicted theoretically already twenty years ago~\cite{huberman1994,sergeev1998,mahan2009,zhangh2013}. 
Although the exact nature of the coupling between the conduction electrons and the substrate phonons 
remains under investigation~\cite{Lu2016}, 
experiments report  conductance values  
in the range $100$ to $1000$~MW/m$^2$/K~\cite{Hopkins2009,guo2012}, in agreement with theoretical predictions. 
One consequence recently emphasized by Lombard et {\em al.}~\cite{Lombard2015} 
is that thermal transport at metal-non metal interfaces is enhanced when such electron-phonon conductance is present.} 
 



Whereas several theoretical and experimental studies have characterized 
heat generation under pulsed illumination of GNPs~\cite{Plech2004,Richardson2006,Juve2009,Baffou2011,Hashimoto2012}, 
the case of CSNPs has received much less attention so far.
Exceptions are Refs.~\cite{Meng2017,Chen2018}, which are, however, restricted to  static conditions only.
%
\modif{
Recent computational studies also investigated the photoacoustic response of silica coated gold nanospheres~\cite{Kumar2018,Shahbazi2019}. 
However, these works consider only one temperature to describe heat transfer in the  core-shell nanoparticle, 
an approximation which is reasonable for nanosecond laser pulses, 
but is insufficient for femto and picosecond pulses.  
Under these conditions, as we will show, electron-phonon processes 
play the leading role in transferring heat to the nanoparticle environment.
}

In this work, we study theoretically transient heat transfer around gold-silica CSNPs illuminated by pulsed laser.
We consider a multi-temperature model, in which each energy carrier is assigned a temperature field. 
Our calculations demonstrate that, at a constant value of the laser power, 
core-shell nanoparticles may heat up the surrounding water faster than gold nanoparticles having the same diameter. 
This effect is significant for thin silica shell and very short laser durations. 
Of prime importance is the role of the interfacial electron substrate conductance which permits rapid heating of the silica shell, 
eventually accelerating heat transfer to water. 
%
The remainder of this article is organized as follows. 
Section.~\ref{sec:model} introduces the model, 
for which an analytic approach  is developed in Sec.~\ref{sec:analytic}. 
Numerical results are presented and discussed in  Sec.~\ref{sec:num}.  
A summary and perspectives are given in Sec.~\ref{sec:conclusion}.

\section{Model}
\label{sec:model}

\modif{
Our model is akin to a two-temperature model. 
We first present the governing equations 
before discussing in details the underlying assumptions and range of validity of this approach.}

\subsection{Governing equations}

\begin{figure}[htbp]
\includegraphics[width=8.5cm]{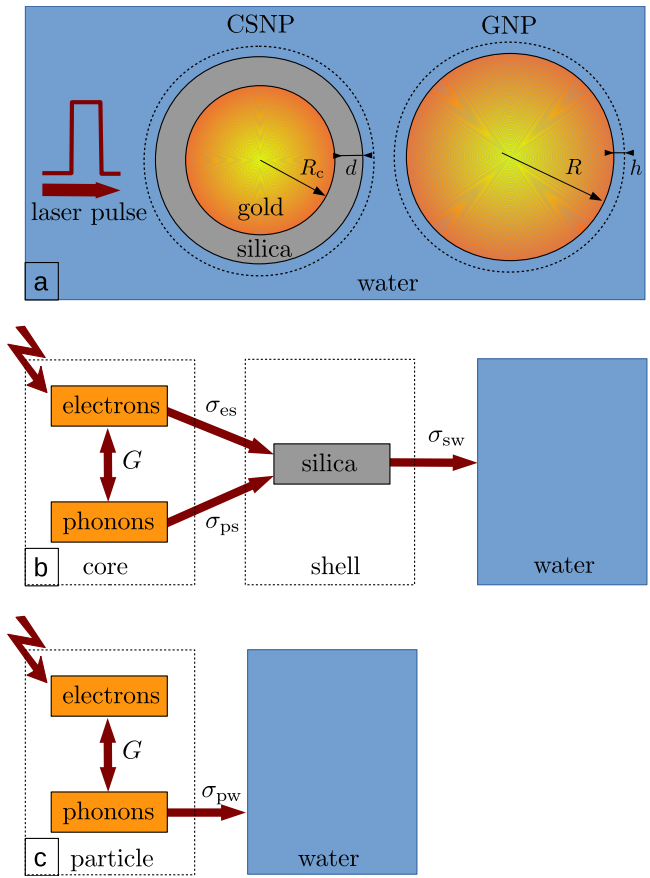}  
\caption{Heat transfer from a core-shell nanoparticle (CSNP) and gold nanoparticle (GNP) heated by a laser pulse.   
$G$~is the coupling constant between electrons and phonons in the metal. 
$\siges$, $\sigps$, $\sigsw$ and $\sigpw$ are the interfacial conductances at the 
electron/silica, phonon/silica, silica/water and phonon/water interfaces respectively. } 
\label{fig:schematic}
\end{figure}

The system we investigate is a core-shell nanoparticle, 
with metallic core and non-metallic shell (Fig.~\ref{fig:schematic}.a).  
Though our description is general, 
we will focus in particular on the gold-silica combination, which has been investigated experimentally~\cite{Chen2011}.
The nanoparticle is immersed in water and submitted to a laser pulsed illumination.  
Figure~\ref{fig:schematic}.b recapitulates the various energy fluxes at play during the relaxation process. 
The light energy  is instantaneously absorbed by the free electrons in the metal causing a rapid increase in their temperature. 
This excess energy  is then transferred to the shell by two different channels.
The first  is direct  and occurs at the core/shell interface through the electron-phonon thermal conductance~$\siges$. 
The second channel is indirect: 
it involves electron-phonon energy exchange inside the core, as governed by the coupling factor~$G$, 
and then  phonon-phonon thermal coupling  at the core/shell interface, quantified by the conductance~$\sigps$. 
With the heat diffusing in the shell, 
a thermal flux  proportional to the phonon-water conductance~$\sigsw$ is induced across the shell/water interface,
allowing the energy to diffuse in the infinite water reservoir. 
Through our focus is on CSNP, 
we will use throughout the study the bare gold nanoparticle as a reference point. 
The heat transfer channels relevant in this case are depicted in Fig.~\ref{fig:schematic}.c: 
while electrons and phonons thermally equilibrate in the metal, 
a thermal flux is induced across the metal/water interface through the interfacial phonon-phonon conductance~$\sigpw$. 

The thermal processes are \modif{supposed to be} governed by the following set of equations
\begin{align}
 \Vc \ce \frac{\partial \Te(t)}{\partial t}&=-\Vc G \left[\Te-\Tp \right]-  \nonumber  \\
                                         & \Sc \siges \left[ \Te-\Ts(\Rc,t)\right ] +\Plaser(t),   \label{eq:VarTe}\\
 \Vc \cp \frac{\partial \Tp(t)}{\partial t}&=\Vc G\left[\Te-\Tp\right]  \nonumber  \\
                                         &  - \Sc \sigps \left[\Tp - \Ts(\Rc,t)\right], \label{eq:VarTp}\\
 \cs \frac{\partial \Ts(r,t)}{\partial t}&=\ks {\nabla}^2 \Ts,  \label{eq:VarTs}\\
 \cw \frac{\partial \Tw(r,t)}{\partial t}&=\kw {\nabla}^2 \Tw.  \label{eq:VarTw}
\end{align}
Here, the index $m\in \{\mathrm{e},\mathrm{p}, \mathrm{s},\mathrm{w} \}$ indicates the component~\footnote{In this order: electrons, metallic phonons, shell phonons and water.} and 
$T_m$, $c_m$, and $k_m$ are respectively  
the temperature, heat capacity  and thermal conductivities of component~$m$.  
The thermal boundary conductance at the interface between component~$m$ and $n$ is indicated by $\sigma_{mn}$. 
While $r$ is the radial distance,  
$\Rc$, $\Sc$ and $\Vc$ are the radius, surface and volume of the core domain. 
$R=\Rc+d$, $S$ and $V$ are the same quantities for the entire nanoparticle, 
including the shell of thickness~$d$, 
whose volume  is $\Vs=V-\Vc$. 
As regards the  power $\Plaser(t)$ received by the electrons, 
we consider for simplicity a square pulse 
$\Plaser(t) = P \Theta(\tpulse-t)$, with $\bar{P}$ the average laser power, $\Theta$ the Heaviside function  and $\tpulse$ the pulse duration.
\modif{Note that the average power $\bar{P}$ is simply related to the laser fluence $F$ and nanoparticle absorption cross-section $\sigma_{\rm abs}(\lambda)$, through~:
  $\bar{P}=\sigma_{\rm abs}(\lambda) F/\tpulse$, where $\sigma_{\rm abs}$ may be described by Mie-theory~\cite{Bohren1983,Baffou2011} for the spherical nanoparticles of interest here.}

In addition to Eqs.~\eqref{eq:VarTe}-\eqref{eq:VarTw}  describing the temporal evolution of temperatures, 
the continuity of the heat flux across an interface implies the 
following boundary conditions at the core/shell interface $r=\Rc$ and the shell/water interface~$r=R$: 
\begin{align}
   -\ks {\nabla} \Ts(\Rc,t) = &\siges \left[\Te-\Ts(\Rc,t)\right] \nonumber \\
                            + &\sigps \left[\Tp-\Ts(\Rc,t)\right],                              \label{eq:bcR}\\
  -\ks{\nabla} \Ts(R,t)  = &\sigsw \left[\Ts(R,t)-\Tw(R,t)\right],  \nonumber \\
                         = & -\kw{\nabla} \Tw(R,t).                                              \label{eq:Jw_int_}
\end{align}
Finally, as regards initial conditions, 
all components are assumed to be at room temperature $\To$. 
The equation for the GNP can be written in a similar manner. 
For completeness, they are given in Supplementary Material (\SM).  

We have listed in Tabs.~\ref{tab:thermophysical} and \ref{tab:Interface_para} 
the parameters relevant for  gold-silica CSNPs. 
\modif{
Two comments are in order on the value of the interfacial conductances. 
First,  
the thermal conductance between the gold nanoparticle and water depends on the wetting contact angle, 
and increases with the hydrophilicity of the interface, as discussed in~\cite{Ge2006,Shenogina2009,Merabia2016}. 
Here, for the evaluation of $\sigsw$, we considered the nanoparticle/water interface to be hydrophilic. 
This choice is justified by the fact that hydrophilic nanoparticle/water interfaces are preferred experimentally, 
in order to ensure a good dispersion of the nanoparticles and avoid nanoparticle agglomeration, 
which may occur if the interface is hydrophobic.  
Second,  
the interface electron-phonon coupling is  assumed to be local,
while it has been proposed that this coupling may extend a few angstr\"{o}ms in the dielectric~\cite{huberman1994,Lu2016}. 
This latter studies demonstrate that the effect of this spatial extension is 
to slightly decrease the total thermal boundary resistance~\cite{Lu2016}. 
Therefore, taking into account these non local effects would only tend to accelerate the heating of liquid water.  
However, since the exact spatial coupling length in silica is not known, 
we will work with a local electron-phonon coupling.}
If not otherwise mentioned, all properties are taken at a temperature $\To=300\,$K. 
There are two quantities, however, 
for which it is essential that the temperature dependence is taken into account. 
The first is  the electronic heat capacity  $\ce=\gamma \Te$ with $\gamma$ the Sommerfeld's constant of the metal~\cite{KittelBook}. 
For gold, $\gamma= 65.6\,$ J$\,$m$^{-3}\,$K$^{-2}$~\cite{Hopkins2009}. 
The second is the interfacial electron-phonon conductance at the core/shell interface, 
which can be written as 
\begin{align}
\siges(\Te)=A+B\Te,
\label{eq:sigesTe}
\end{align}
with the constant 
$A = 96.1\,$MW$\,$m$^{-2}\,$K$^{-1}$   and $B=0.18\,$MW$\,$m$^{-2}\,$K$^{-2}$~\cite{Hopkins2009}. 

\begin{table}[htbp]
	  \centering
	  \begin{tabular}{ c c c c c l }
	    \hline
	    \hline
	     \ quantity\  & \ electron\          & \ phonon\     & \ silica\        & \ \ water\ \   &unit\    \\
	    \hline
	    $k$           & $320.0$      & $3.0$     & $1.37$       & $0.606$ &  W$\,$m$^{-1}\,$K$^{-1}$     \\ 
	    $c$           & $0.0197$     & $2.35$    & $1.01 $      & $4.11$  &  MJ$\,$m$^{-3}\,$K$^{-1}$   \\
	    $\dif$        & $16243$      & $1.28$    & $1.36 $      & $0.15$  &   nm$^2\,$ps$^{-1}$          \\
	    \hline
	    \hline
	  \end{tabular}
	  \caption{ Thermophysical parameters at $\To=300\,$K for the components of the considered system:   
	  thermal conductivity~$k$, heat capacity~$c$, and diffusivity~$\dif=k/c$.}
	  \label{tab:thermophysical}
\end{table}

\begin{table}[htbp]
	\centering
	\begin{tabular}{l r l  l}
	    \hline
	    \hline
	    quantity\ \ \ \ \ \  & value        	& unit & Ref. \\ 
	   \hline
	    $G$               & $2.5 \times 10^{10}$ &  MW$\,$m$^{-3}\,$K$^{-1}$ & \cite{Lin2008}      \\
	    $\siges(\To)$     & $152$                &  MW$\,$m$^{-2}\,$K$^{-1}$ & \cite{Hopkins2009}  \\
	    $\siges(10\,\To)$ & $636$                &  \ \ \ \ \ \ \ -          & \cite{Hopkins2009}  \\
	    $\sigps$          & $141$                &  \ \ \ \ \ \ \ -          & \cite{Lombard2015}  \\
	    $\sigsw$          & $150$                &  \ \ \ \ \ \ \ -          & \cite{Hu2009}        \\
	    $\sigpw$          & $139$                &  \ \ \ \ \ \ \ -          & \cite{Ge2006,Shenogina2009,Merabia2009b} \\    
	    \hline
	    \hline
	\end{tabular}
	\caption{Thermal exchange coefficients in bulk gold and at the interfaces between the various components.} 
	\label{tab:Interface_para}
\end{table}

\subsection{Discussion}

\modif{A complete treatment of the heat transfers in our system 
should be based on a  Boltzmann transport equation~(BTE) 
with rates that account for all possible scattering channels~\cite{Qiu1993b,Wang2016}. 
However, BTE calculations are computationally demanding, 
in particular when different phonon modes should be described. 
The method we consider here is only approximate, 
but since the computational resources required are modest, 
it opens the way to an efficient exploration of 
the parameter space defined by the nanoparticle and pulse features.  
As an additional benefit, 
the approach is amenable to an analytical treatment, 
an insight which is difficult to obtain with the BTE. 

The governing equations given above define 
a four-temperature model for the CSNP-water system, 
which is a straightforward extension of the two-temperature model~(TTM)~\cite{Kaganov1957,Anisimov1974}. 
The key point of the TTM is 
to describe the nonequilibrium dynamics of electrons and phonons, 
by assuming coupled rate equations for the different energy carriers. 
Predictions of the TTM were partly confirmed by 
early experimental observations of non equilibrium between electrons and phonons in metal 
using time domain thermoreflectance as reported by Eesley~\cite{Eesley1983}.
The  TTM is based on a number of standard but simplifying assumptions.
Because we apply this framework for particles having size below $100\,$nm 
and for laser excitation  of very short time scales (down to $1\,$ps),  
it is important to discuss the expected range of validity. 
We discuss in turn the three main assumptions underlying our governing equations:\\ 
A1. Fast  electron thermalization. \\ 
A2. Diffusive or ballistic transport  of the energy carriers. \\ 
A3. Bulk near-equilibrium  values for the transport coefficients.  \\

A1 {\em Fast electron thermalization.} 
Electron thermalization depends on the rate of both electron-photon and electron-electron processes. 
The electron-photon interactions
first bring the the valence electrons~\footnote{$6$s in case of gold.} to a high energy state~\cite{Arbouet2004}. 
The initial electron distribution is a nascent (as photoexcited) distribution 
which differs from the Fermi-Dirac distribution 
$f_{\rm FD}(\epsilon,\mu(\Te))=1/\exp((\epsilon-\mu(\Te))/k_\mathrm{B}\Te)$~\cite{Fann1992a,Fann1992b}. 
Indeed, at very short time scales 
excited electrons are not "dressed" by electron-electron interactions and thus they may not be viewed as quasiparticles. 
Is it only after sufficient electron-electron collisions have occured 
that the electron density may by described by a Fermi-Dirac distribution 
with a well-defined local electronic temperature~$\Te$. 
The corresponding electron-electron  scattering time $\tauee$ 
may be inferred from liquid Fermi theory~\cite{Ashcroft1976,Nozieres1966} 
and depends on the excess energy of the electron $\delta \epsilon=\epsilon-\epsilon_F$ with respect to the Fermi level $\epsilon_F$: 
$\tauee= \tau_0(\epsilon_F/\delta \epsilon)^2$ in the RPA approximation~\cite{Nozieres1966,Fann1992a}, 
where $\tau_0=128/(\pi^2\sqrt{3}\omega_p)$ and $\omega_p$ the plasma frequency.
This scattering time describes the relaxation time of an electron due to its interaction with a reservoir of thermalized electrons. 
The thermalization time of the electrons, however, 
is a collective phenomenon occuring on a time scale~$\tau$, 
which may be calculated using Boltzmann transport equations 
of the electron-electron scattering process characterized by the time~$\tauee$. 
Monte Carlo simulations demonstrated  that the relaxation time~$\tau$ generally decreases with the laser fluence~\cite{Mueller2013}.
Note that the sensitivity of the electron relaxation time on the laser fluence 
explains why early experimental investigations, occuring at low excitation levels, 
reported nonthermal character of the electron distribution~\cite{Fann1992a,Fann1992b,Tas1993,Sun1993,Groeneveld1995,Schoenlein1987}.  
At higher excitations, the electron cooling down is relatively well described 
by the two-temperature model~\cite{Wang1994,Wang2012,Mueller2013,Giri2015b}.

For the specific case of gold, and for electron temperatures higher than $800\,$K, 
the relaxation time computed from Monte Carlo simulations~\cite{Mueller2013}
turns out to be shorter than $1\,$ps, 
and in this work we will concentrate on this latter regime.  
For femtosecond pulses, this implies that the  incident laser power 
is above a minimum value that is typically a few watts. 
In those conditions, 
the fast electron thermalization underlying TTM is a reasonable approximation.

A2 {\em Diffusive or ballistic transport of the energy carriers.} 
As demonstrated by first-principle calculations~\cite{Jain2016}, 
the vast majority of phonon modes in gold have a mean free path smaller than $10\,$nm~\footnote{Note 
that the  phonon mean free path we  refer to here originates in phonon-phonon scattering processes.}. 
Because the gold nanostructures of interest here are significantly larger, 
the assumption of diffusive transport for  gold phonon is well justified 
(some alternatives have been proposed in~\cite{Chen2005}). 
In contrast, 
it has been shown experimentally that the electron mean free path in gold is $\delta=100\,$nm~\cite{Qiu1993a,Hohlfeld1999,Bonn2000,Lejman2014}, 
which is  above the typical size of the nanoparticle  considered in this work. 
In view of this quasi-ballistic nature of the transport, 
we  assumed that the  electronic temperature~$\Te$ is uniform all across the metallic region (see Eq.~\eqref{eq:VarTe}). 
Furthermore, 
the spatial dependence of gold phonon temperature was also neglected (see Eq.~\eqref{eq:VarTp}), 
assuming in effect a flat profile. 
As discussed in Appendix~\ref{ap:diffusion}, 
this hypothesis is numerically validated in the nanostructures under investigation.   
One convenient implication 
is that the radial symmetry is conserved at all time. 

As regards amorphous silica, recent molecular dynamics~(MD) simulations demonstrated that, in this system, 
Fourier's law is valid down to the nanometer scale~\cite{Larkin2014}, 
justifying our assumption of diffusive transport in silica shell. 
Similarly, heat transport in liquid water can also be modelled as diffusive. 
Recent MD simulations~\cite{Rajabpour2019} have indeed showed that, 
in the vicinity of a metallic nanoparticle, 
the local thermal conductivity of water may be described by bulk water conductivity, 
except in the immediate vicinity~($2\,$nm) of the nanoparticle and at very short time scales ($t<5$ ps).



A3 {\em Bulk  near-equilibrium transport coefficients. }
Given that the gold nanoparticles considered below may have a radius as low as $20\,$nm, 
one may wonder  whether we should consider the electron-phonon coupling constant of bulk gold. 
Early experimental investigations aimed at determining the electron-phonon coupling constant 
in metallic thin films~\cite{Schoenlein1987,Brorson1990,Elsayed-Ali1991,Elsayed-Ali1987,Norris2003} 
were prompted by theoretical predictions of the thermal relaxation in metals by Allen~\cite{Allen1987}. 
In the case of nanoparticles, Arbouet et {\em al.} demonstrated for diameters above $10\,$nm, 
the intrinsic electron-phonon interaction is well accounted for by the bulk constant~\cite{Arbouet2003}. 
This being said, 
we should also discuss the fact that the electron-phonon coupling constant~$G$ may change with the electronic temperature~$\Te$
\cite{Chen2005,Lin2008,Giri2015b,Giri2015c}. 
Actually, this dependence  starts to be significant at high temperatures $\Te>5000\,$K for gold, 
and in this study we will concentrate on lower energy excitations. 
For the sake of completeness, we have performed TTM simulations in CSNPs  
taking into account the temperature dependence of both the electron-phonon coupling constant, and the electronic heat capacity. 
As detailed in Appendix~\ref{ap:non_linear_Gep}, 
the results indicate that these non-linear effects have little impact in the regime considered. 
  
In conclusion, 
the TTM  framework may provide a sound description of thermal transport across gold-silica nanoparticles, 
whose gold region has spatial dimension larger than $10\,$nm, 
and irradiated  by  laser pulses of moderately high fluence, 
such that the maximal electronic temperature $\Te$ is between $800$ and $5000\,$K. 
For femtosecond pulses, this corresponds typically to incident laser powers between $2$ and $100\,$W. 
Note finally that since such a power is below the threshold for water cavitation~\cite{Lombard2017}, 
we discard all phenomena related to mass transfer or bubble formation. 
}

\section{Analytical approach}
\label{sec:analytic}

Here we examine by analytical means to what extent heat relaxation dynamics 
differ in CNSP and GNP. 
To do so, 
we consider the model defined by Eqs.~\eqref{eq:VarTe}-\eqref{eq:VarTw}, 
but for the sake of tractability, 
we add  the following  approximations: 
i) heat diffusion is instantaneous, implying that the temperature is uniform within each domain and 
that in particular, the shell thickness does not play any role, 
ii) all coefficients are constant, 
iii) the heating of electrons is instantaneous,  
and 
iv) the water remains at initial temperature $\To$. 
%
The system of equations is now linear and the solution can be obtained in terms of Laplace transforms with respect to time. 
Transforming back to the time domain to obtain explicit expressions for the temperature $\{T_m(t)\}$ 
is not possible analytically in the general case 
but it is straightforward to do so once parameters are given their numerical values. 
Because $\ce$ and $\siges$ both depend heavily on temperature, 
we will consider two limiting cases, 
where they take a low and high value respectively. 
We use $\ce(T)$ and $\siges(T)$, 
with $T$ fixed to $\To$ and $10\,\To$ in case A and B respectively. 

\begin{figure}[htbp]
\includegraphics[width=8cm]{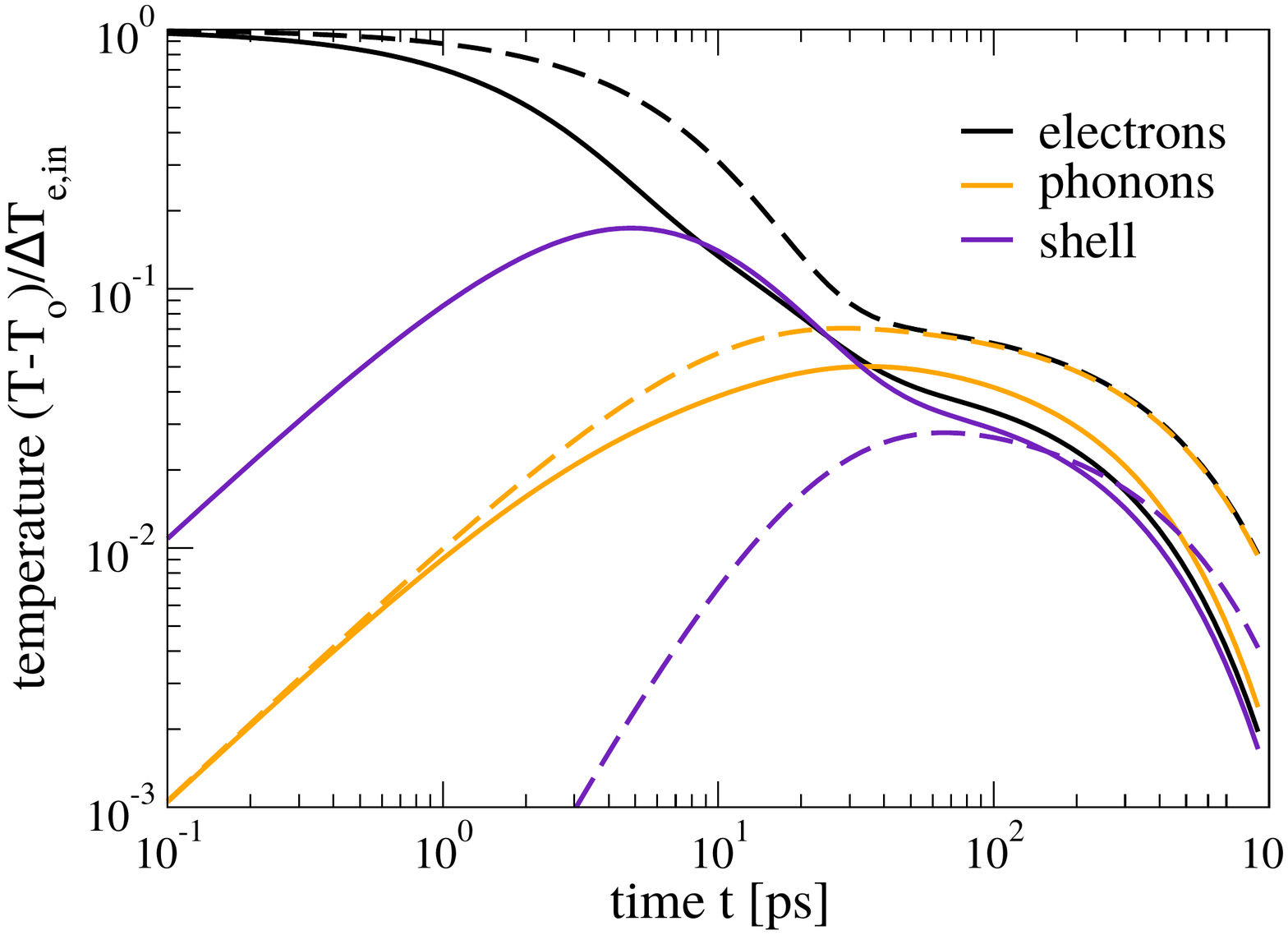}
\caption{Time evolution of the temperatures for the components of a CSNP, 
with and without interfacial electron-shell conductance (continuous and dashed line respectively), 
as predicted by the analytical model. 
The initial electron temperature shift is unity.  
} 
\label{fig:anTt}
\end{figure}

The time evolution of temperatures is shown in Fig.~\ref{fig:anTt}, 
where we compare the result with and without the electron-silica conductance. 
The most striking difference is visible at short time. 
With~$\siges$, 
the silica shell temperature~$\Ts$ is well above that of gold phonons $\Tp$, and reaches a maximum before $10\,$ps. 
Without~$\siges$, 
$\Ts$ remains largely below~$\Tp$, and the maximum is attained at a much later time, around  $100\,$ps.  
The acceleration of energy transfer in the presence of electron-silica conductance 
is particularly visible in the electron temperature, 
whose decrease is shifted to earlier time over most of the time range. 
Note finally that with~$\siges$  there is a time interval somewhere between $1$ and $100\,$ps where the silica shell has the highest temperature. 
Because all those features are common to case A and B, 
one can expect they are relevant in the full model as well, 
which is confirmed by comparison with the numerical results (see Fig.~\ref{fig:profiles} below). 
Overall, 
this suggests that the analytical solution is able to capture at least the qualitative features.  

We now characterize the short-time behavior of heat transfer. 
Because water is assumed to remain at temperature~$\To$, 
we  consider the total heat flux going out of the particle. 
It turns out to grow linearly at early time 
\begin{align}
J/\delTein =  \kappa t  + \bigO(t^2), 
\label{eq:Jout}
\end{align}
where $\delTein$ is the initial shift in electron temperature. 
The prefactor is found to be 
\begin{align}
 \kappacs      &=  \frac{12 \pi (\Rc+d)^2}{\Rc \left[ (1+d/\Rc)^3-1\right]} \frac{\siges \sigsw}{\cs},      \label{eq:kappa} \\
 \kappabare  &= 4 \pi R^2 \frac{ \sigpw G}{\cp},                                                      \label{eq:kappabare}
\end{align}
for the core-shell and gold nanoparticle respectively. 
Except for the smallest particles with $R \lesssim 15\,$nm,  
the $\kappacs/\kappabare$  ratio  computed for particles of same total radius $R=\Rc+d$
is approximately 2 and 10, in case A and B respectively. 
The  heat flux at short time is thus significantly enhanced with core-shell nanoparticles. 
The thin-shell limit $\epsilon=d/\Rc \rightarrow 0$ is the most interesting 
and leads to the very compact expression for the ratio
\begin{align}
\frac{\kappacs}{\kappabare}   &=  \frac{\cp \siges \sigsw}{G \cs \sigpw } \left( \frac{1}{\epsilon} -1 \right) + \bigO(\epsilon),
\end{align}
which is independent of radius, 
and where the numerical prefactor is 0.38 and 1.6 in case A and B respectively. 
In the latter case, an order of magnitude enhancement in flux 
is expected for the typical value~$\epsilon \simeq 0.1$, 
suggesting that a thin shell CSNP is clearly to be preferred over a GNP.

One way to quantify the relaxation rate of heat transfer
is to consider the time  required to heat up the water in the vicinity of the particle. 
Accordingly, we propose an estimate for the time~$\taus$ 
needed to reach temperature~$\To+\delTw$ in a water layer of thickness~$h$. 
We assume that the  heat released by the particle, as estimated by integration of Eq.~\eqref{eq:Jout}, 
serves only to heat up homogeneously the water layer, without diffusing farther, giving
\begin{align}
\frac{\taus^2 \delTein}{2 \kappa}   &=  \delTw \cw \frac{4 \pi}{3} \left[ (R+h)^3 - R^3  \right]. 
\label{eq:tstardef}
\end{align}
For $h=2\,$nm, 
we get $\tauscs=66$ and $32 \sqrt{\delTw/\delTein}\,$ps in cases~A and~B respectively.  
This line of reasoning is applicable 
only if the time  $h^2/\difw=26\,$ps needed for heat diffusion in water over a distance~$h$, 
is comparable to~$\tauscs$, 
which is the case as long as $\delTw/\delTein$ remains below unity.

\begin{figure}[thbp]
\includegraphics[width=7cm]{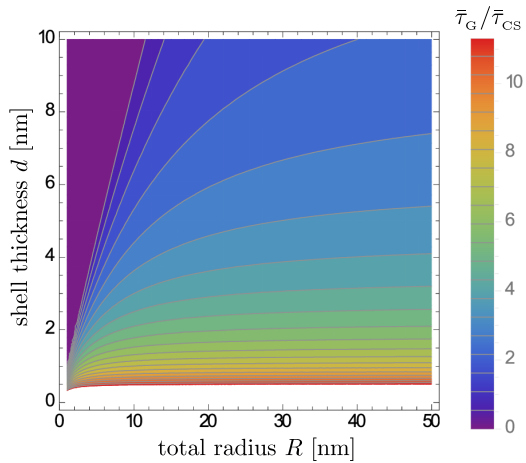}
\caption{The ratio of relaxation time $\tausbare/\tauscs$  for CSNP and GNP, 
as a function of total radius~$R$ and shell thickness~$d$ for case B. 
The result for case A is quite similar with value typically twice smaller. } 
\label{fig:tstaratio}
\end{figure}

A very compact expression for $\tauscs$ is obtained 
if we take again the thin-shell limit and expand at lowest order in~$h/\Rc$: 
\begin{align}
\tauscs   &=  \sqrt{\frac{ d \cs }{\siges} \frac{  h \cw }{ \sigsw}   \frac{\delTw}{\delTein} }, 
\label{eq:tstar}
\end{align} 
where one recognizes the geometric average of the two relaxation times associated with 
interfacial conductance $\siges$ and $\sigsw$. 
Numerically, $\tauscs=43$ and $21 \sqrt{\delTw/\delTein}\,$ps in cases~A and B respectively. 
With $\delTein \simeq 10^3$  and $\delTw$ in the range $1-100\,$K, 
$\tauscs$ falls in the range $0.6-13\,$ps, 
which is close to the heating time that will be found numerically below, 
at least for intermediate and high laser powers.  
In the low power regime, near the threshold, 
heat presumably has time to diffuse in water~\cite{Lombard2017} and Eq.~\eqref{eq:tstar} is not expected to apply. 
Finally, we can compare $\tauscs$ for the core-shell and gold nanoparticles. 
From Eqs.~\eqref{eq:kappa}, \eqref{eq:kappabare} and \eqref{eq:tstardef}, 
one sees that the ratio $\tausbare/\tauscs$ is controlled by $\sqrt{\kappabare/\kappacs}$.  
As visible in Fig.~\ref{fig:tstaratio},  
the enhancement is more pronounced for thin shells. 
For a typical case where $R=45\,$nm and $d=5\,$nm, 
one gets for case A and B a ratio $\tausbare/\tauscs=1.6$ and $3.3$ respectively. 

To conclude this section,  
our analytical approach suggests that compared to GNP, 
CNSP with thin shells may induce heat transfer that are two or three times faster.  
As we show in Appendix~\ref{ap:theolongtime}, 
the presence of a silica shell may also decrease the longest relaxation time in the system. 
To probe the intermediate regime which is most relevant to applications 
and obtain a complete picture, taking into account non-linearities, 
we now turn to numerical resolution.

\section{Numerical results and discussion}
\label{sec:num}

We  present in this section our numerical results on the heat transfer around illuminated nanoparticles, 
with a focus on the comparison between CSNP and GNP. 
The simulation method is based on a finite difference scheme.
In contrast to the analytical treatment of Sec.~\ref{sec:analytic}, 
where only the short and long time behaviors were accessible, 
we have now access to the entire relaxation process and 
we can assess the heating efficiency 
with criteria that are tailored to a specific application. 
For a broad range of photothermal applications, 
the primary goal is to heat up the immediate vicinity of the particle, 
so as to transfer heat to the environment. 
This is true not only for photoacoustics~\cite{Chen2011}, but also for photothermal bubble generation~\cite{Lombard2014}. 
%
Therefore, we choose as our central quantity the heating time $\theating$ needed 
to reach a given temperature $T_0+\delT$ at a distance $\dwater$ from the surface of the nanoparticle.  
Throughout the study, 
we fix $\dwater=2\,$nm, whereas $\delT$ will be varied in a large range. 

As a complementary measure of efficiency,  
we will also consider the ratio 
\begin{align}
 \eta = \frac{E_{\dwater}(\theating)}{E_{\rm laser}(\theating)}. 
\end{align}
Here $E_{\dwater}(t)$ is the energy stored  in water within a distance~$\dwater$  from the particle surface 
and $E_{\rm laser}(t)$ is the total energy supplied by the laser. 
Both quantities are time dependent and we consider their ratio at time~$\theating$ when the desired temperature level $\delT$ is reached.

Our default parameters are as follows.  
The initial temperature is $\To=300\,$K. 
If not otherwise mentioned, 
the total radius of the particle is $R=45\,$nm,  the silica shell has thickness~$d=5\,$nm 
and the laser pulse duration is $\tpulse=10\,$fs.

\subsection{CSNP vs GNP}

\begin{figure*}[thbp]
\centering
\includegraphics[height=5.5cm]{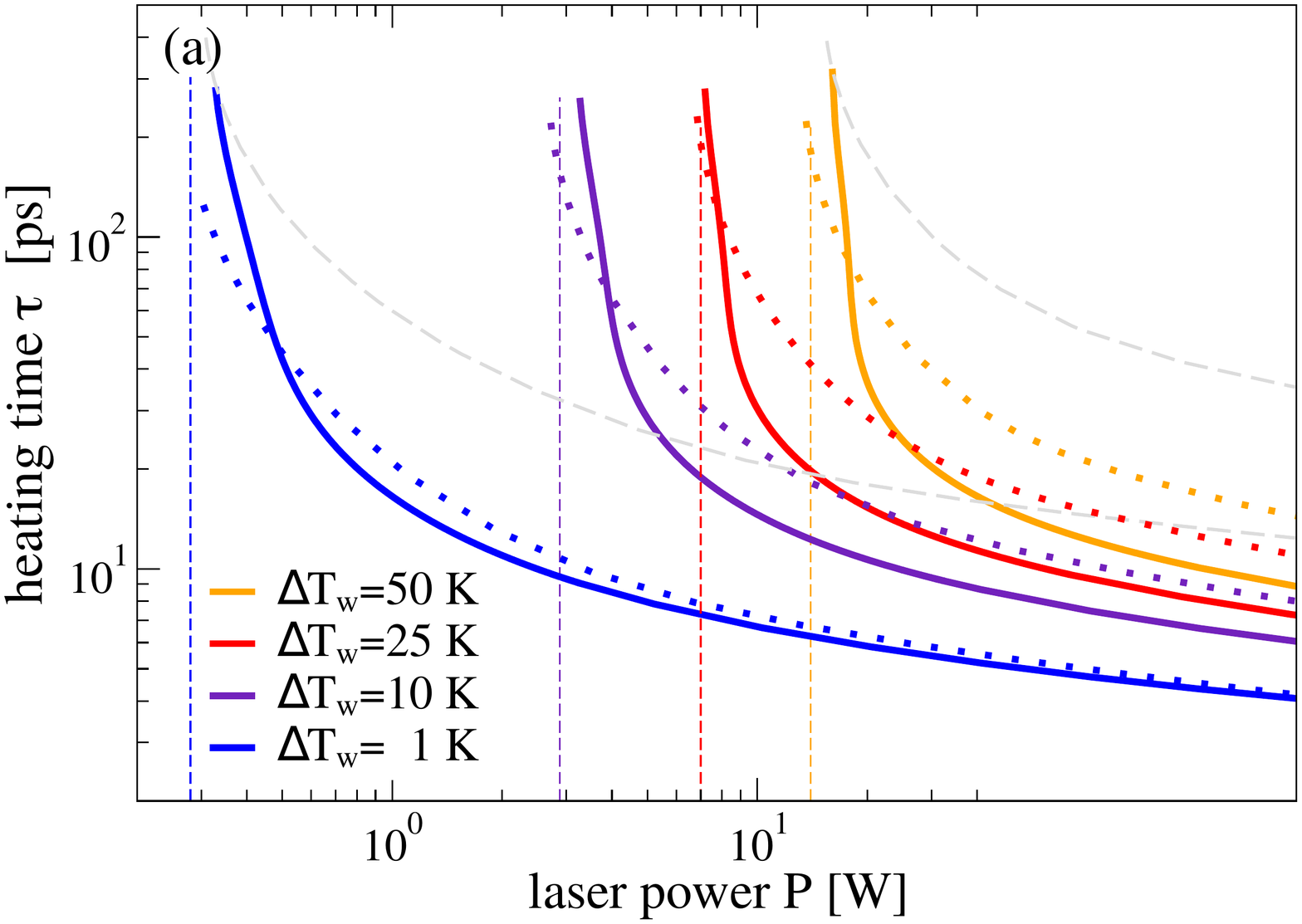} \hspace*{1cm}
\includegraphics[height=5.5cm]{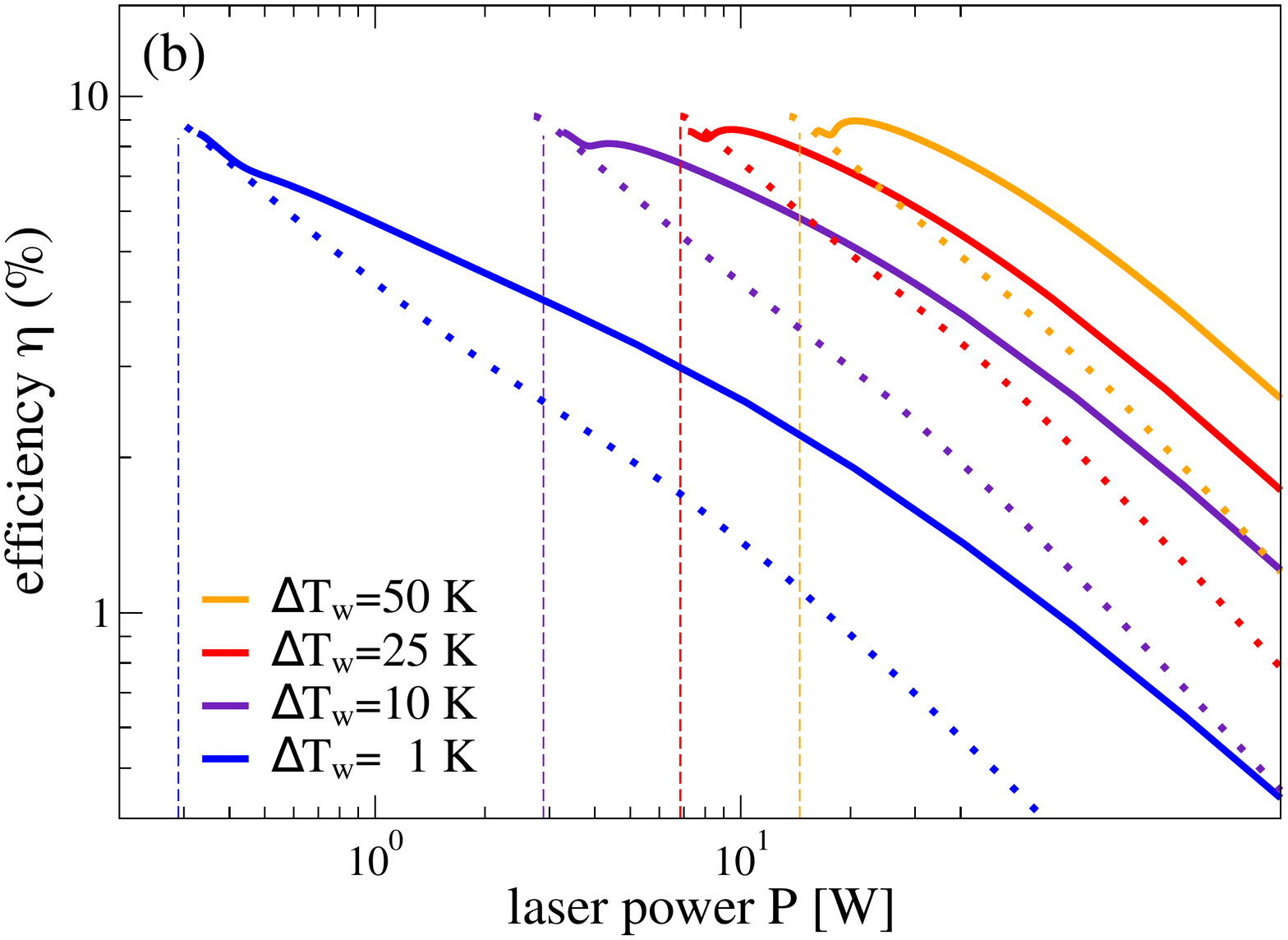}
\caption{Heating time~$\theating$~(left) and efficiency~$\eta$~(right) as a function of laser power $P$, 
for  various temperature shifts $\delT$ of the water shell. 
Results are included for  CSNP  (solid lines) and GNP  (dotted lines). 
In both cases, the total radius is $R=45\,$nm. 
The dashed gray lines show  the heating time for CSNP 
when the electron-silica channel is disabled ($\siges=0$),  
for $\delT=1\,$K (left curve) and $\delT=50\,$K (right curve).  
Vertical lines indicate the threshold power $\Pth$.}  
\label{fig:HeatingTimeDT}
\end{figure*}

Figure~\ref{fig:HeatingTimeDT}.a shows the heating time $\theating$ 
needed to reach a desired temperature shift~$\delT$ ranging from~$1$ to $50\,$K. 
We first note that $\theating$ is defined only if 
the laser power is above a threshold value  $\Pth$, 
which is proportional to the temperature shift. 
As can be expected, 
the heating time is a decreasing function of the laser power, 
with a steep drop  in the vicinity of the threshold.  
Our main observation is that, away from the threshold, 
the CSNP heats up the water faster than GNP. 
The trend is most pronounced at high heating level. 
For instance, with $\delT=50\,$K and a laser power \modif{$P=50\,$W}, 
the heating time is \modif{almost twice shorter} with a CSNP. 
The decrease of the heating time due to the presence of a silica shell is of the same order of magnitude as what is reported experimentally~\cite{Hu2003}.
The difference in the heating time between the two types of particles
vanishes at low heating level: keeping \modif{$P=50\,$W,} it is \modif{only~$5\%$} at  $\delT=\,1$K.  
This trend can be understood if we recall that the  interfacial electron-silica conductance $\siges$ 
depends heavily on temperature, as specified in Eq.~\eqref{eq:sigesTe}. 
As detailed below, 
the enhancement of heat transfer is thus more significant at high temperature.

Core-shell and gold nanoparticles also differ in their efficiency, as visible in  Fig.~\ref{fig:HeatingTimeDT}.b. 
Strikingly, except very close to the threshold, 
the CSNPs  always perform better.    
For instance, with laser power  \modif{$P=50\,$W}  and $\delT=1\,$K,  
there is more than a  two-fold factor in efficiency, 
with \modif{$\eta=1.2\%$ and $0.5\%$} for CSNP and GNP respectively. 
Note that whatever the temperature shift, the efficiency never exceeds 10\% 
and is only a fraction of a percent at the highest power laser. 
In this case, 
most of the energy brought by the laser pulse either diffuses away or is stored elsewhere than in the water shell. 
Because the heating time is minimum well above the threshold, 
whereas the efficiency is maximal near the threshold, 
an optimal value of laser power should exist, 
if the two criteria are to be optimized simultaneously.

\subsection{Role of the interfacial electron-silica conductance}

\begin{figure*}[thbp]
\centering
\includegraphics[height=5.5cm]{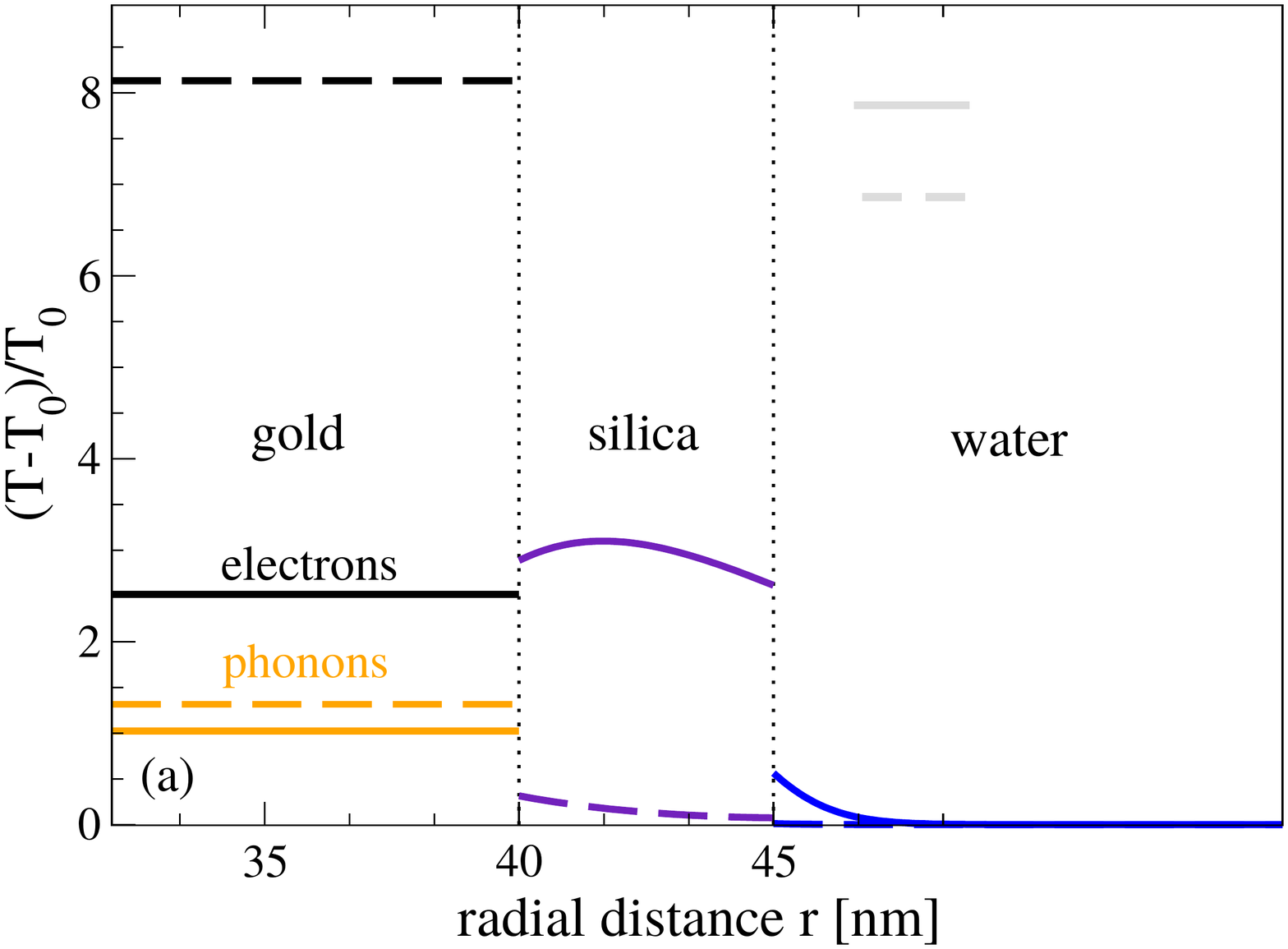} \hspace*{1cm}
\includegraphics[height=5.5cm]{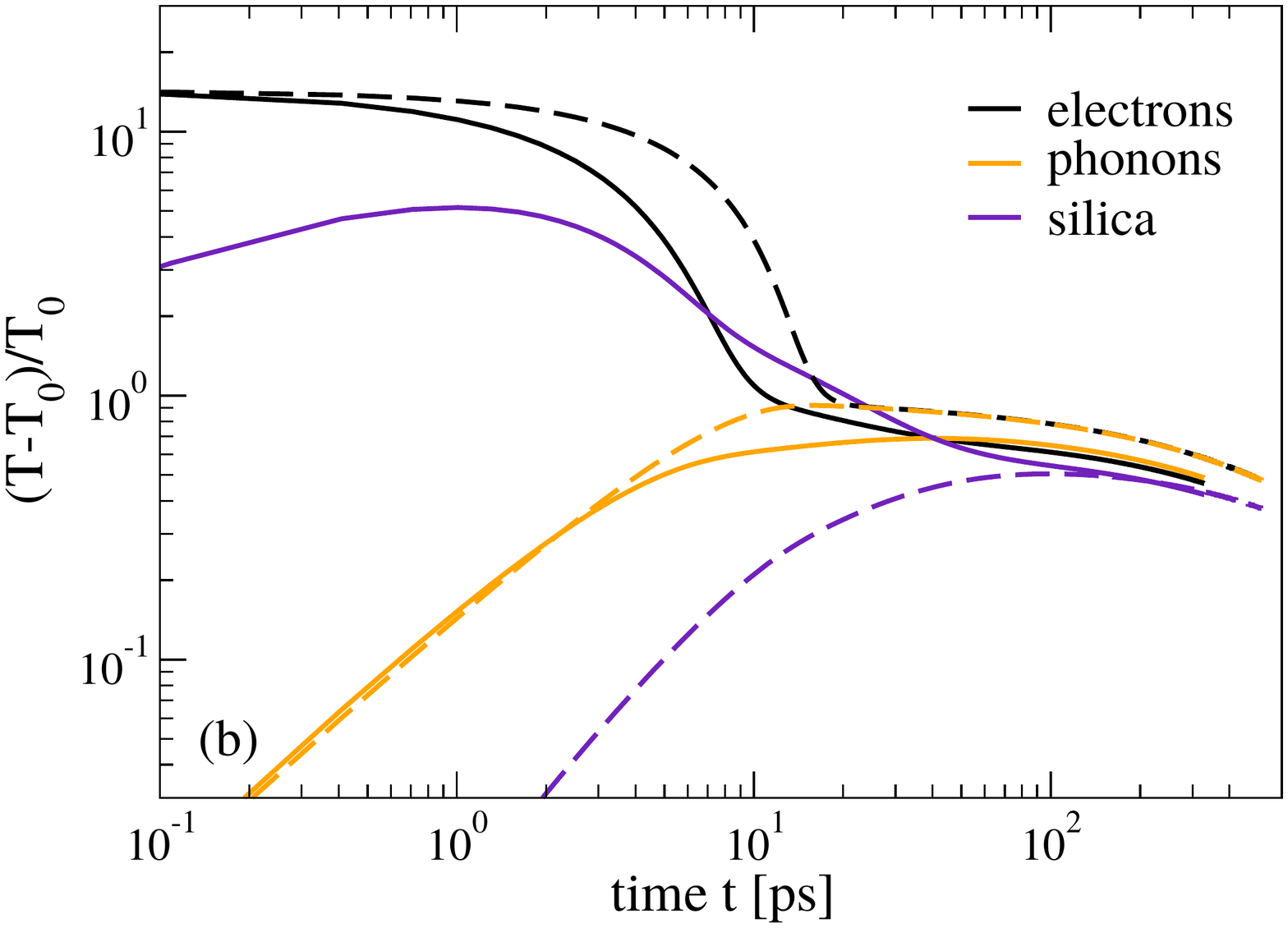}
\caption{
Temperature within CSNP with and without electron-silica conductance $\siges$ (continuous and dashed lines respectively).  
(a) 
Local temperature for each domain at time $t=10\,$ps. 
The laser power is $P=13.9\,$W, close to the threshold.  
(b)
Time evolution of the mean temperature of each component. 
}  
\label{fig:profiles}
\end{figure*}

CSNPs may heat up the neighbouring water shell faster and more efficiently than GNPs. 
If however, the electron-silica coupling is disabled ($\siges=0)$, 
this conclusion no longer holds, 
as illustrated in Fig.~\ref{fig:HeatingTimeDT}.a. with the dashed lines.  
The heating time would become larger for CSNPs. 
To better appreciate the role of $\siges$ on the heating dynamics,  
we plot in Fig.~\ref{fig:profiles} the temperature profile within the CSNP and its time evolution, 
with and without $\siges$. 
Whereas the phonon temperature is rather insensitive to the presence of this channel, 
both the electron and silica temperatures are strongly affected. 
For instance, at time  $t=10\,$ps shown in Fig.~\ref{fig:profiles}.a, 
the electron temperature is four times lower, 
whereas the silica temperature is higher by an order of magnitude. 
Surprisingly, 
the silica component can be the hottest component in the system. 
Shown in Fig.~\ref{fig:profiles}.b  are the time evolutions of the different temperatures in the system. 
The electrons are the first to reach a maximum  at the end of the illumination by the laser power (not shown). 
Around a time of $1\,$ps, 
the silica  in turn reaches a maximum at $\Ts \simeq 1800\,$K, 
well above the phonon temperature at this stage, 
a phenomenon already noticed with the analytical model.
The picture is completely modified without $\siges$. 
The maximum in silica temperature is much lower ($\Ts \simeq 450\,$K) 
and occurs at a much latter time ($10^2\,$ps). 
Thus, 
the  acceleration of the energy transfer 
that was already present in the simplified analytical model 
is fully confirmed in the numerical results.  

\subsection{Influence of particle size and shell thickness}

The ability of a particle to store energy is related to its geometrical and its thermophysical parameters, 
notably the thermal diffusivity~$\dif$ and heat capacity~$c$, 
which are fixed once materials are chosen. 
On the other hand, 
both the size of the metallic core and the thickness of the silica shell can be tailored at will.  
In what follows, we investigate their influence on the energy transfer.

\begin{figure*}[thbp]
\centering
\includegraphics[height=5.5cm]{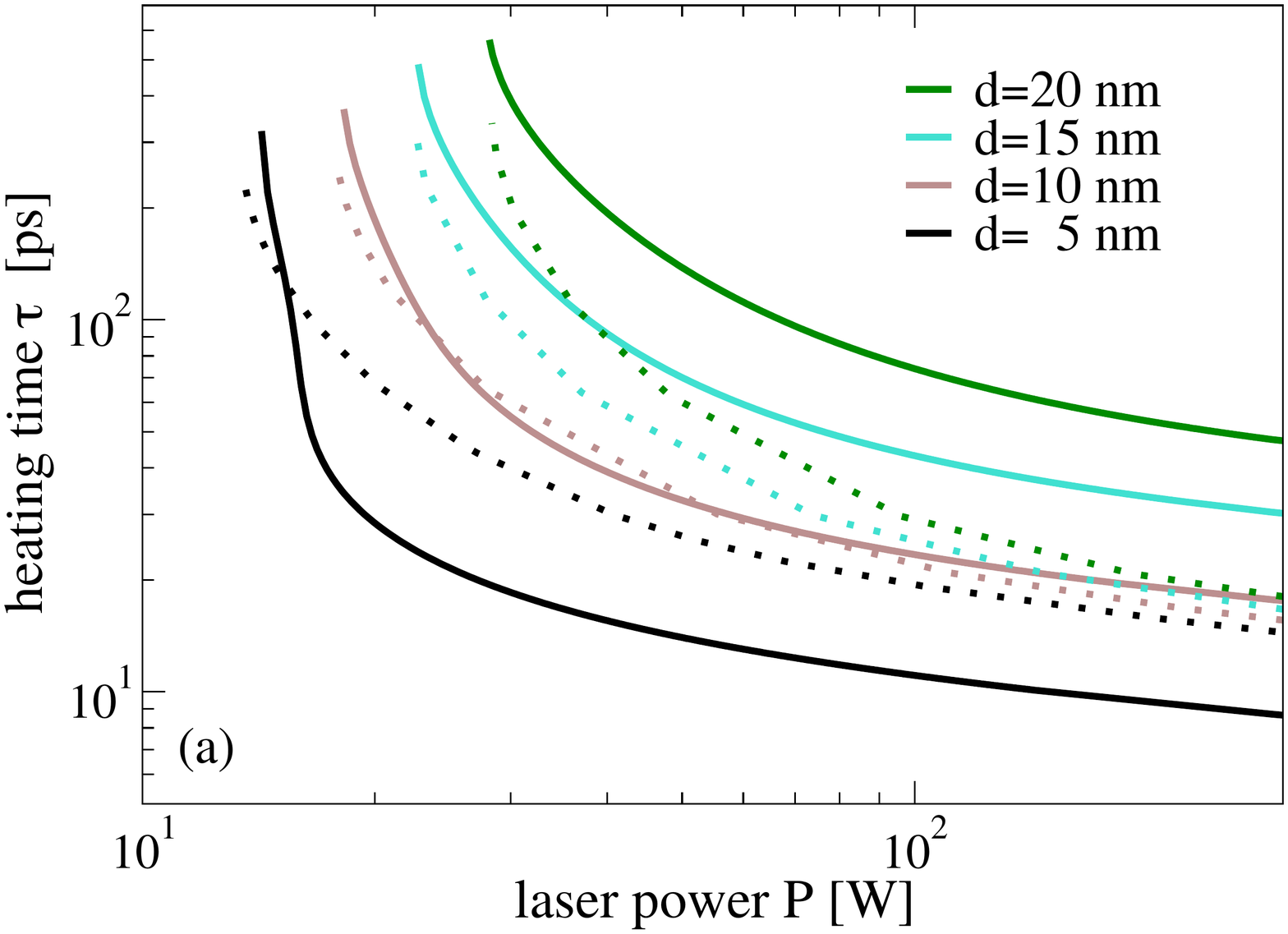} \hspace*{1cm}
\includegraphics[height=5.5cm]{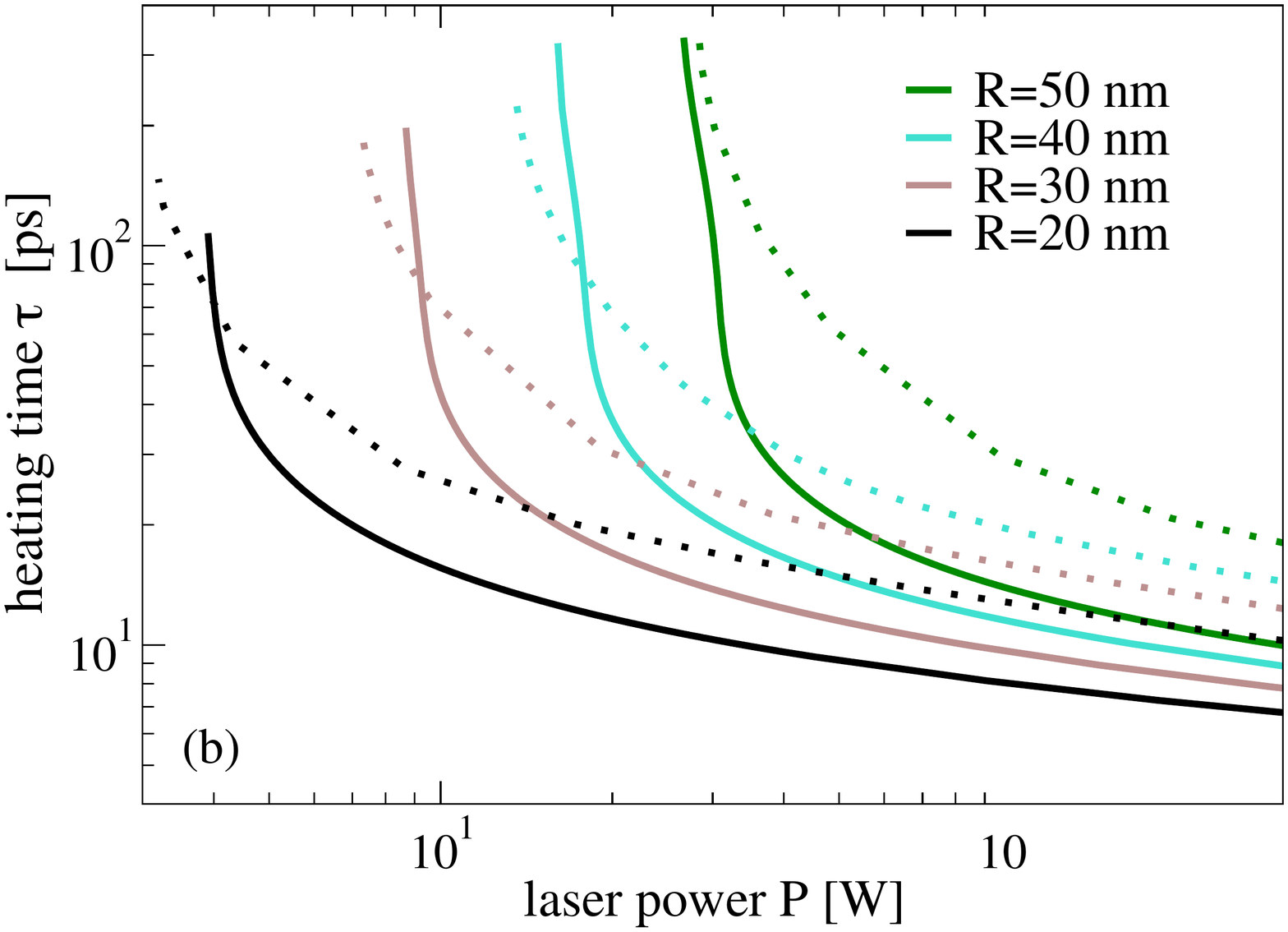}
\caption{Influence of geometric properties of CSNP on heating time. 
(a) The shell thickness~$d$ is varied while keeping a core radius~$\Rc=40\,$nm.  
(b) The total radius~$R$ is varied, while keeping a shell thickness~$d=5\,$nm. 
The temperature shift is $\delT=50\,$K. 
The dotted lines indicate the results for a GNP of same total radius. 
}  
\label{fig:size}
\end{figure*}

Figure~\ref{fig:size} illustrates the effect of changing the shell thickness~$d$. 
Here, the gold core radius is fixed to $\Rc=40\,$nm, 
and the shell thickness increases from $5$ to $20\,$nm.  
The temperature shift is \modif{$\delT=50\,$K}.  
For comparison,  
the heating time for a GNP with same total radius $R=\Rc+d$ is also reported. 
A feature common to both types of particles when increasing size, 
is the shift in the threshold laser power, 
which reflects the fact that, for larger particles,  more  energy is needed to reach the same heating level in the surrounding water shell.
Now, above the threshold,   
CSNP and GNP display different behaviors. 
When the laser power~$P$ exceeds $100\,$W, 
the heating time of GNPs varies only slightly with their size. 
In contrast, 
the heating time of CSNP increases considerably with shell thickness. 
Such a trend can be expected since the thermal conductivity of silica is low compared with that of metal, 
and heat transfer is strongly limited by heat diffusion in the silica shell (see \SM\  for an illustration). 
As a result, 
a CSNP with the thickest shell $d=20\,$nm yields a heating time 
which is almost an order of magnitude longer than that of a  GNP of equal radius, 
whereas the thinnest shell leads to a relaxation that is still faster. 
It is thus clear that thin shells favor a fast heat relaxation. 

The core particle size can also impact the heating performance. 
Fixing the silica shell thickness at $d=5\,$nm,  
the core radius is now varied from 20 to $50\,$nm. 
The heating time is shown in Fig.~\ref{fig:size}.b 
and appears to decrease with the core size.  
%
It turns out that small core-shell particles are beneficial in two ways. 
First, the heating time decreases with the nanoparticle size.  
Second, CSNPs perform better then GNPs due to the electron/phonon interfacial coupling. 
If one seeks fast heating in the water shell,  
the optimal nanoparticle appears to be a small CSNPs with thin shell.  


\subsection{Influence of pulse duration}

Results so far involved a laser pulse of duration $10\,$fs.  
In this subsection, we investigate how the dynamics of heat transfer is affected by a much longer pulse, 
in the picosecond or nanosecond range. 
Though the mechanisms are the same, 
the heating time  and threshold power exhibit drastic changes.

\begin{figure*}[thbp]
\centering
\includegraphics[height=5.5cm]{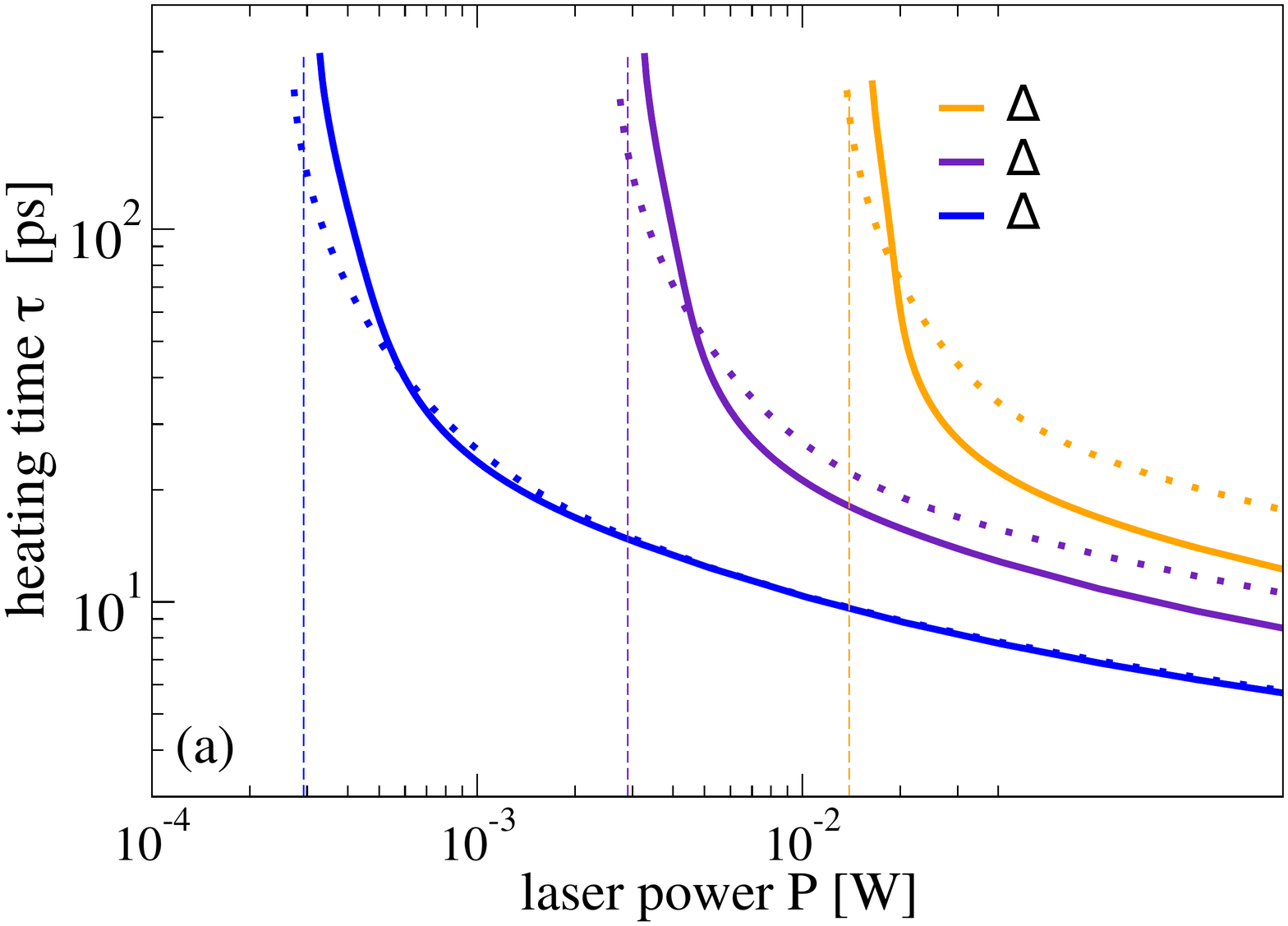} \hspace*{1cm}
\includegraphics[height=5.5cm]{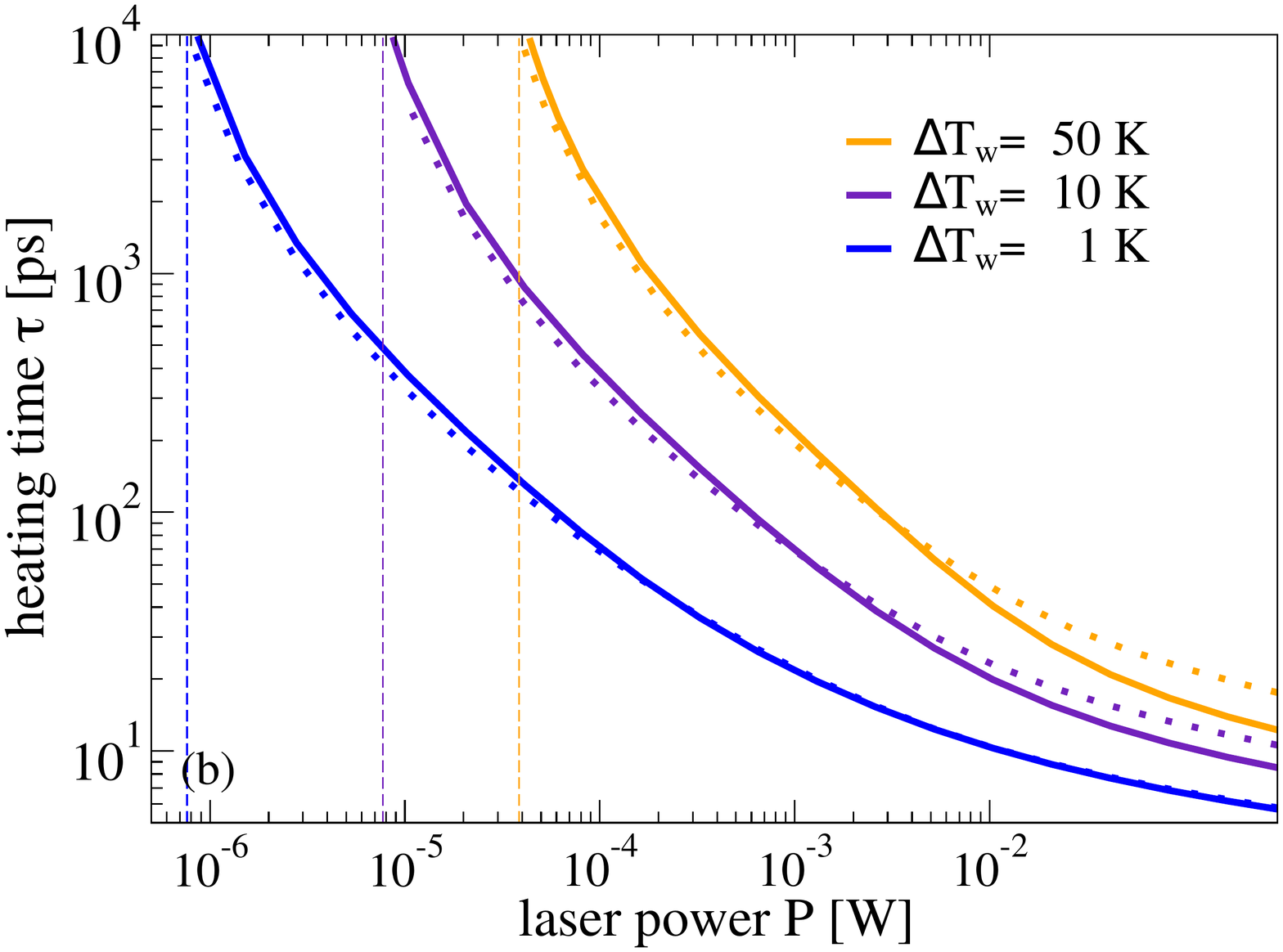}
\caption{Influence of the laser pulse duration on the heating time: 
$\tpulse=10\,$ps  in panel~(a) and $10\,$ns in panel~(b).   
The water temperature shift is $\delT$. 
Results for CSNP and GNP are shown in continuous and dashed lines respectively. 
Here, $\Rc=40\,$nm and~$d=5\,$nm.}
\label{fig:longpulse}
\end{figure*}

The heating time for a pulse duration $\tpulse=10\,$ps is shown in Fig.~\ref{fig:longpulse}.a~\modif{
\footnote{
\modif{For pulses of duration $10\,$ps and $10\,$ns,  
the maximal electron temperature remains below $5000\,$K as long as the laser power is below $10^{-1}\,$W. 
Besides, from an extrapolation of Fig.~11 in~\cite{Mueller2013}, 
the electron thermalization time in gold at $300\, $K  is around $5\,$ps
which does not exceed the pulse duration. 
The fast thermalization assumption is thus acceptable. }}}  
When comparing with the short-pulse result shown Fig.~\ref{fig:HeatingTimeDT}, 
the typical values for $\theating$  are similar 
but the threshold power decreases by  three orders of magnitude.  
For instance, for \modif{ $\delT=50\,$K, 
we have $\Pth = 0.014\,$W rather than $14\,$W previously.}
This trend continues for much longer pulses. 
Shown in Fig.~\ref{fig:longpulse}.b is the heating time for  a laser duration  $\tpulse = 10\,$ns. 
The threshold powers now fall in the range  $10^{-4}-10^{-6}\,$W depending on the temperature level. 
Choosing again  \modif{$\delT=50\,$K}, 
one sees that  $\Pth$ has dropped one more time by \modif{more than two} orders of magnitudes, 
reaching now $\Pth \simeq 4\times 10^{-5}\,$W.
As regards the heating time, 
they now extend over a much larger range, up to $10^4\,$ps.  
In the vicinity of the threshold, 
they are  typically on the order of the pulse time, 
implying that all the laser energy is needed before the heating criterion can be fulfilled. 

A remarkable observation here is that for the longest pulse time, 
the heating time of CSNP and  GNP are almost the same in most of the laser power range,  
except for the highest values $P \gtrsim 10^{-2}\,$W.  
This applies whatever the desired temperature shift. 
To understand this result, 
we first note that the typical time for the thermal energy  to diffuse within the silica shell 
is $d^2/\difs \simeq 18\,$ps, 
which is much lower than the pulse duration $\tpulse=10\,$ns. 
Therefore, the heat transfer is not limited by equilibration within the silica domain. 
To get further insight, 
we have looked at the temperature evolution within CSNP and GNP 
(see Fig.~S2 in \SM). 
Except for the very early time before $1\,$ps, 
the electron, phonon and water temperatures are identical in the two types of particles.  
Thus, 
in contrast to short pulses for which the three particles temperature are far from equilibrated, 
long laser pulses leave sufficient time for thermalization of the different components within the particle. 
The heat relaxation is then limited by the transfer to water and diffusion therein. 
Given that the gold-water conductance and silica-water conductance 
have close values (see Tab.~\ref{tab:Interface_para}), 
one can expect a very similar dynamics for the CSNP and GNP. 

\section{Conclusion}
\label{sec:conclusion}

To summarize, we presented a complete study of heat transfer around illuminated gold-silica core-shell nanoparticles immersed in water. 
We described the energy transfer by analyzing a simplified model and 
then solving numerically the heat transfer equations.  
Our model takes into account the thermal boundary conductance at the particle/water and core/shell interfaces, 
and highlights the role of the interfacial electron-phonon coupling. 
Throughout the study, we compare results for a CSNP and a GNP  having the same size. 
To assess the heating dynamics, 
we chose  as a quantitative criterion the time needed to increase the temperature 
of a thin narrow shell of water surrounding the particle.  
Our results show that in many situations the CSNPs enable heat transfers that are faster than GNP. 
We identified optimal conditions to minimize the heating time:
a short-laser pulse and a thin silica layer.  

\modif{We focused exclusively on the silica-coated gold nanoparticles, 
but it is worth discussing whether our approach and conclusions may extend to other systems.
The effect that we outlined here relies on the electron-phonon interfacial coupling~$\siges$. 
This coefficient has been measured experimentally for gold/dielectric interfaces~\cite{Hopkins2009,guo2012,Giri2015a}.
but remains essentially unknown for other metals. 
Therefore, it would be extremely valuable 
to obtain from  time domain thermoreflectance or  first principle calculations, 
the interface conductance in other cases, such as  silver or aluminium. 
A further question  is to ascertain whether the assumptions inherent to the TTM framework 
still apply in the case of other systems. 
There are several clues already available. 
First, we note that gold has a relatively long  electron thermalization time as compared to other metals, 
as shown by Boltzmann transport calculations~\cite{Mueller2013}. 
Second,  
gold has a rather small electron mean free path~\cite{Qiu1993a}. 
Finally, first principle calculations demonstrate that for gold, aluminium and silver, 
the distribution of phonon mean free paths essentially vanish above $10\,$nm. 
Therefore, we conclude that for metals other than gold, 
our TTM treatment should hold under comparable and possibly broader conditions of laser heating. 
}

Our analysis may help in interpreting two types of experimental investigations. The first involves time-resolved spectroscopy experiments~\cite{Hu2003}, which showed that gold-silica core-shell nanoparticles in water display faster heat dissipation than gold nanoparticles. Second, our analysis confirmed the scenario put forward in recent photoacoustic studies~\cite{Chen2011,Chen2012} 
which interpreted the enhanced photoacoustic signal displayed by gold silica nanoparticles, as
stemming from the improved heat transfer efficiency to the environment due to the presence of a silica shell. 
%
As we showed in this study, the key process here is the coupling between the metal electrons and silica phonons. 
If the silica shell is thin enough, the mechanism    
may offer optimized conditions for heat transfer from the gold core to the nanoparticle environment.
\modif{
As regards the nonlinear photoacoustic response of CSNPs, 
it may be analysed either by using the analytical approach of Calasso et {\em al.}~\cite{Calasso2001} in the linear regime, 
orby solving the finite difference time domain (FDTD) numerical scheme of Prost et {\em al.}~\cite{Prost2015}.
The frontier between the linear and non-linear regime corresponds to a laser fluence $F \simeq 7\,$ mJ/cm$^2$, 
which for CSNP irradiated by femtosecond laser pulses with a wavelength in the vicinity of the surface plasmon resonance corresponds to a power $\Plaser \simeq 20$ W.  
Note that a recent study solved photoacoustic equation for silica-coated nanoparticles based on Fourier heat conduction model~\cite{Shahbazi2019}. 
However, this work does not treat electron-phonon coupling processes, 
and addresses mainly  long laser pulses (in the ns range). 
In contrast, our study demonstrates that for femtosecond and picosecond laser pulses, 
electron-phonon couplings can not be ignored, and yield enhanced thermal response of the nanoparticle environment. 
This would motivate  a full determination of the photoacoustic response of CSNPs following pulsed irradiation 
taking into account thermal diffusion and coupling between {\em all} the energy carriers. }

This study has implications both on the applied and fundamental sides. 
First, it calls for the experimental or theoretical determination of the electron-phonon conductance of different combinations of metal-dielectric interfaces. 
Of special interest will be interfaces displaying high thermal coupling. 
Enhancing locally heat transfer is important for a broad array of applications, 
among which photoacoustic generation  may be the most illustrative example~\cite{Chen2011,Prost2015}. Here, the simplest type of core-shell nanoparticle has been envisaged, but clearly electron/phonon interfacial couplings should play major role in a wide class of heterogeneous particles, including concentrated core-shell multilayers~\cite{Meng2017,Bardhan2010}. Combining the optical response of these structures to heat transfer mediated by the different channels opens an avenue for the optimization of photothermal energy conversion based on metallic dielectric multilayers nanostructures.


\appendix

\section{Longest relaxation time in theoretical model}
\label{ap:theolongtime}

\begin{figure}[thbp]
\includegraphics[width=7.5cm]{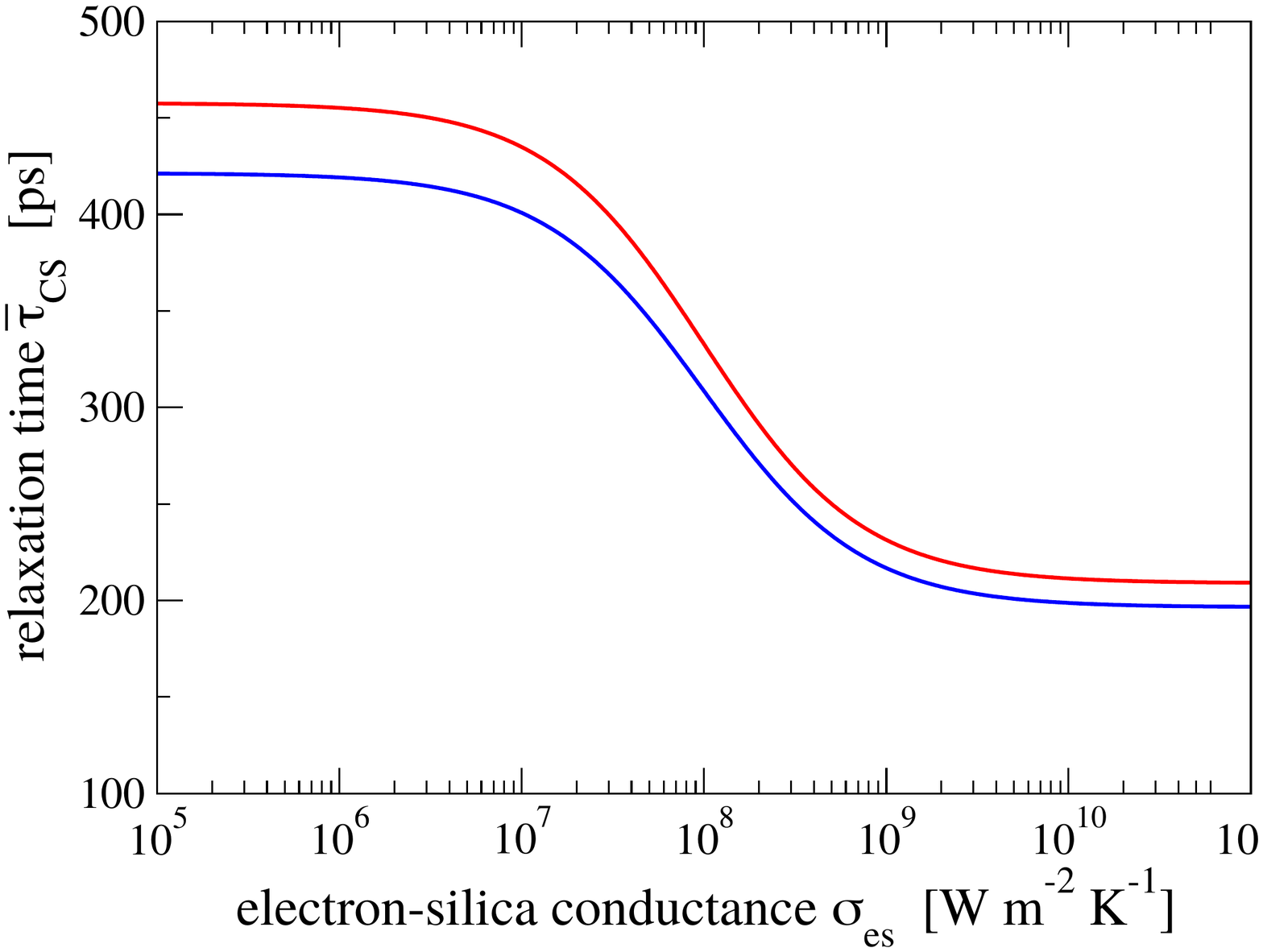}
\caption{ Relaxation time $\taulcs$ as a function of $\siges$, for $\ce$ fixed to  $\ce(\To)$  and $\ce(10\,\To)$
(bottom and top curve respectively).} 
\label{fig:tau-siges}
\end{figure}

Because the water temperature is assumed to remain at its initial value~$\To$, 
the long-time behavior is exponential 
and involves a relaxation time  which we characterize in this section. 
Skipping the derivation, one gets for the core-shell nanoparticle
\begin{align}
 \taulcs =& \frac{1}{3} \frac{G \Rc (\ce+\cp) + 3 \ce \sigps }{G \Rc (\siges+\sigps)+ 3 \siges \sigps}  \nonumber \\
         & + \frac{\Vc (\ce+\cp) + \Vs \cs}{S \sigsw},  
\end{align}
which includes the two limiting cases 
\begin{align}
\siges \longrightarrow \infty,   \quad \taulinf &= \frac{\Vc (\ce+\cp) + \Vs \cs}{S \sigsw}, \\
\siges \longrightarrow 0,    \, \quad \taulo    &=   \frac{\ce}{G} + \frac{\Rc(\ce+\cp)}{3 \sigps} + \taulinf. \label{eq:taulim0}
\end{align}
Here, $\taulinf$ is the relaxation time for a core-shell nanoparticle 
where all components (gold electrons and phonons, silica phonons) 
would be equilibrated at the same temperature. 
$\ce/G$ is a typical time for equilibration between electrons and phonons. 
The second term in the RHS of Eq.~\eqref{eq:taulim0} would be the typical relaxation time 
for the ensemble  ``electrons+phonons'', assuming they have the same temperature. 
To assess the effect of the electron-silica conductance, 
the relaxation time $\taulcs$ is plotted as a function of $\siges$ in Fig.~\ref{fig:tau-siges}. 
Whatever the value chosen for  $\ce$, 
the trend is similar. 
After taking constant value at low  $\siges$, 
the relaxation time decreases to another plateau which is approximately twice lower. 
Note that the two values considered in case A and B 
fall around  the upper end of the range $10^7-10^9$ MW $\,$ m$^{-2}$ $\,$ K$^{-1}$ over which the drop takes place. 
Thus, as previously found with other criteria, 
one also sees in the long-time behavior that 
neglecting  the electron-silica conductance 
would significantly slow down the relaxation in the system.

To compare the case of core-shell and gold nanoparticles, 
we compute the longest relaxation time for the latter and find
\begin{align}
\taulbare & =  \frac{\ce}{G} + \frac{R (\ce+\cp)}{3 \sigpw}, 
\end{align}
where one can recognize the sum of 
the electron-phonon equilibration time, 
and the relaxation time for a bare nanoparticle where electron and phonon are equilibrated. 
This latter equation generalizes the equation ruling the heat dissipation time of a bare nanoparticle with Kapitza resistance~\cite{Wilson2002}, the latter time being obtained in the limit 
$G \rightarrow \infty$ and $\ce \rightarrow 0$.

\begin{figure}[thbp]
\includegraphics[width=7.5cm]{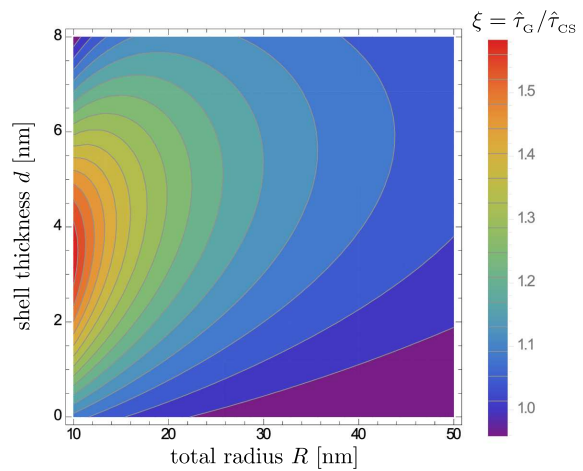}
\caption{ Ratio of the longest relaxation time for CSNP and GNP, 
as a function of total radius~$R$ and shell thickness~$d$, in case B. 
Here the diffusion in silica shell is accounted for in a approximate manner (see text). } 
\label{fig:xi-Rd}
\end{figure}

Introducing the ratio of relaxation times  $\xi  = \taulbare/\taulcs$  
for core-shell and bare particles having the same size $\Rc+d=R$, 
we find that 
the core-shell nanoparticles are more advantageous ($\xi$~above unity)
at small size and thicker silica shells. 
At this point, one should remember that our theoretical model 
assumes infinitely fast diffusion in the silica shell, 
implying that it is only valid for small shell thickness~$d$. 
An adhoc but simple way to account for diffusion is to add to $\taulcs$
the typical time  for diffusion in the silica shell, $d^2/\difs$. 
This is done in Fig.~\ref{fig:xi-Rd}
which show the $\xi$~ratio  as a function of total radius and shell thickness. 
As expected, the optimal shell thickness is now finite. 
For instance, it is around $6\,$nm for a particle of total radius $45\,$nm. 
Note that the enhancement is weaker in this case,  
with $\xi$ ratio mostly below $1.5$.

\section{Effect of non-linear temperature dependence of electron-phonon coupling constant G and electronic heat capacity $\ce$}
\label{ap:non_linear_Gep}

\begin{figure}[htbp]
  \includegraphics[height=5cm]{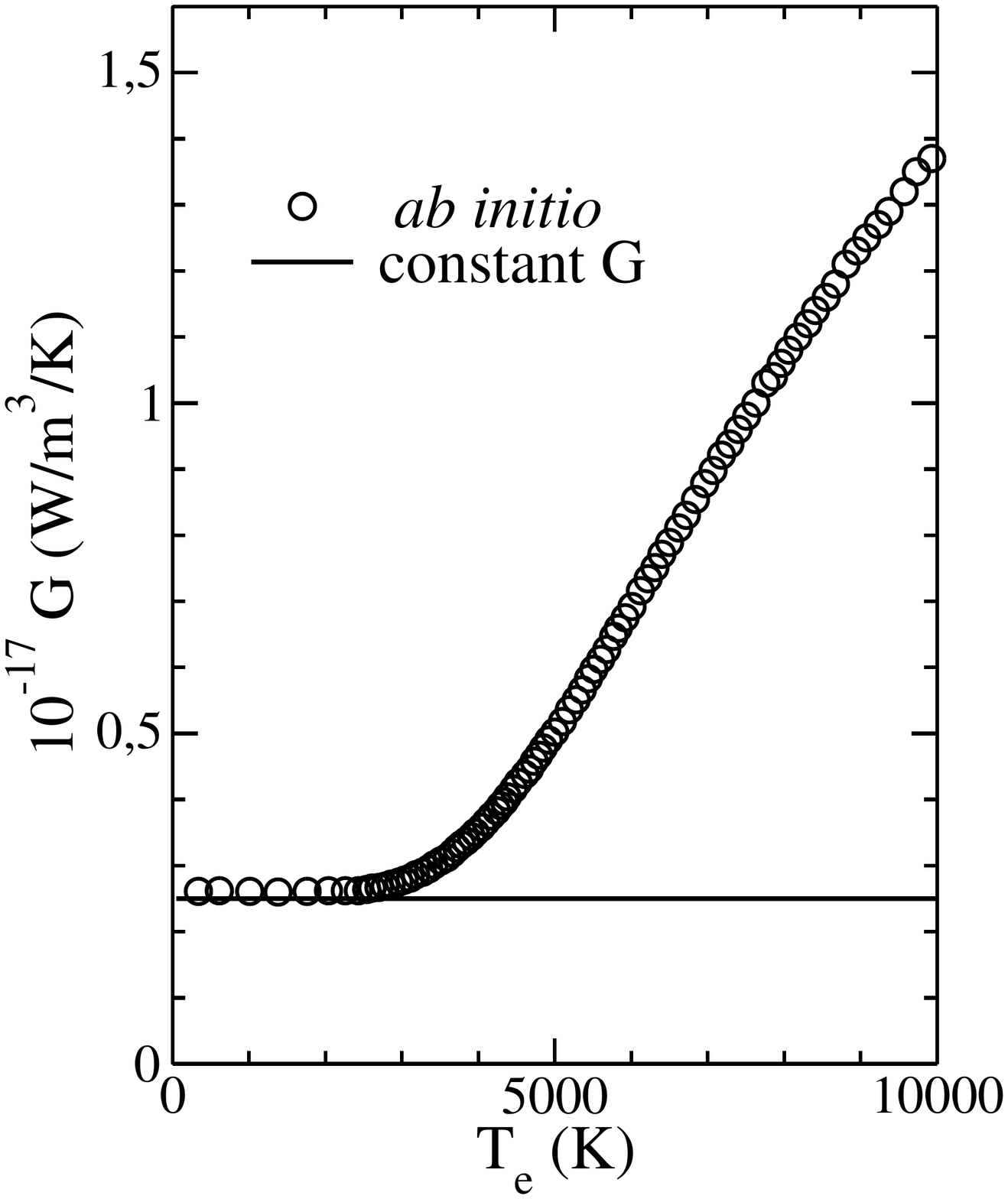} 
  \includegraphics[height=5cm]{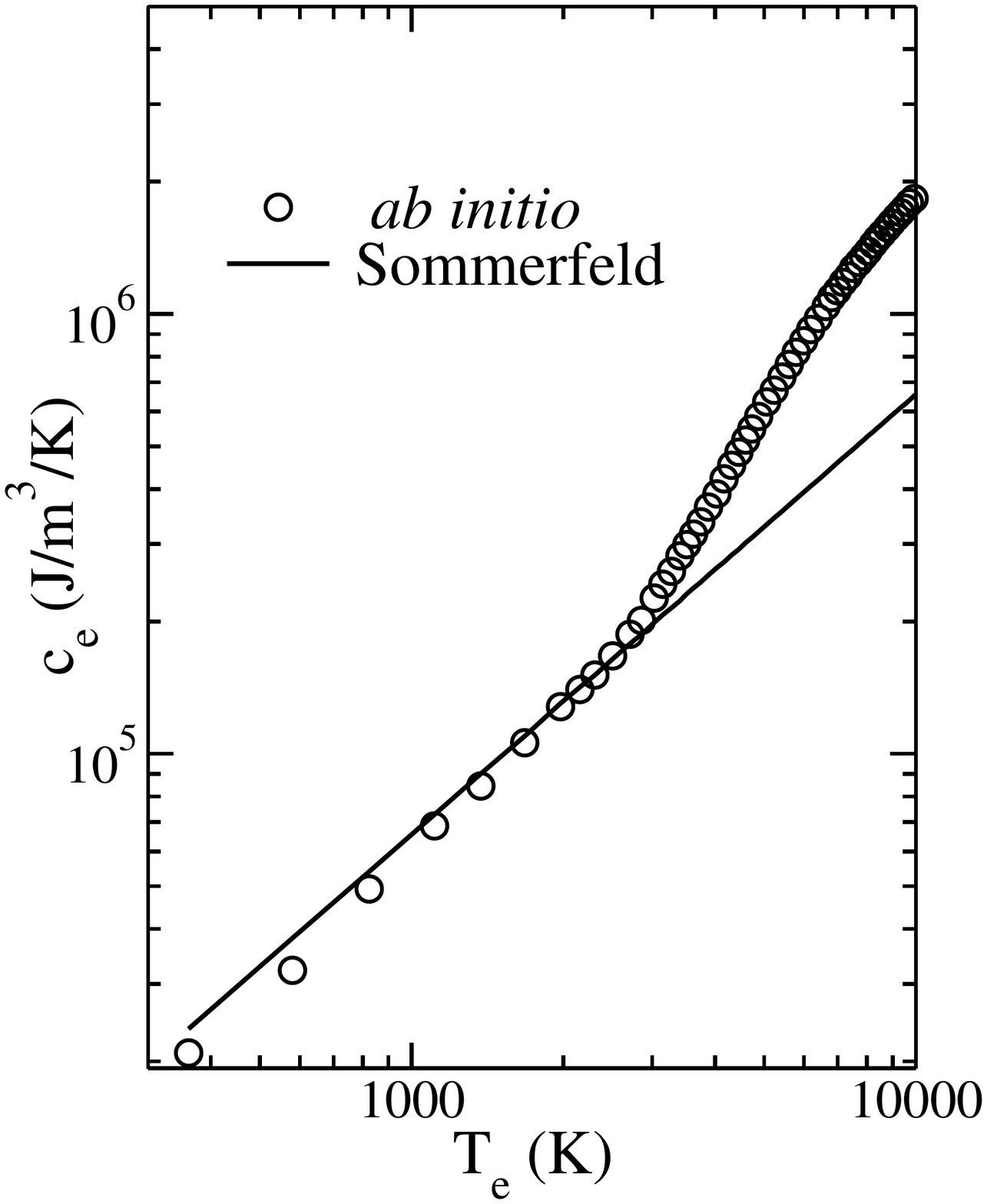}
\caption{
\modif{Electron-phonon coupling $G$~(left) and electronic heat capacity $\ce$~(right) of bulk gold as a function of the electronic temperature. 
The results of first principles {\em ab initio} calculations (symbols)  are compared 
to the near-equilibrium coupling constant and to the Sommerfeld heat capacity~(lines).}}
\label{fig:abinitioGepCe}
\end{figure}

Throughout our study, 
we  used for the electron-phonon coupling constant~$G$ and the electronic heat capacity $\ce$ 
the values  close to equilibrium. 
Under these conditions, $G$ is constant, 
while $\ce=\gamma \Te$ increases with the electron temperature, $\gamma$ being the Sommerfeld constant. 
For electronic temperatures higher than approximately $4000\,$K for gold, 
interband transitions start to play a role,  
both $G$ and~$\ce$ deviate from their near equilibrium values,
and become non-linear functions of $\Te$, as shown by first-principle {\em ab initio} calculations~\cite{Lin2008}.  
As visible in Figure~\ref{fig:abinitioGepCe}, 
the deviations from the near-equilibrium predictions for $G$ and $\ce$ occur above an electronic temperature $\Te>3000\,$K.

\begin{figure}[htbp]
\includegraphics[width=7.5cm]{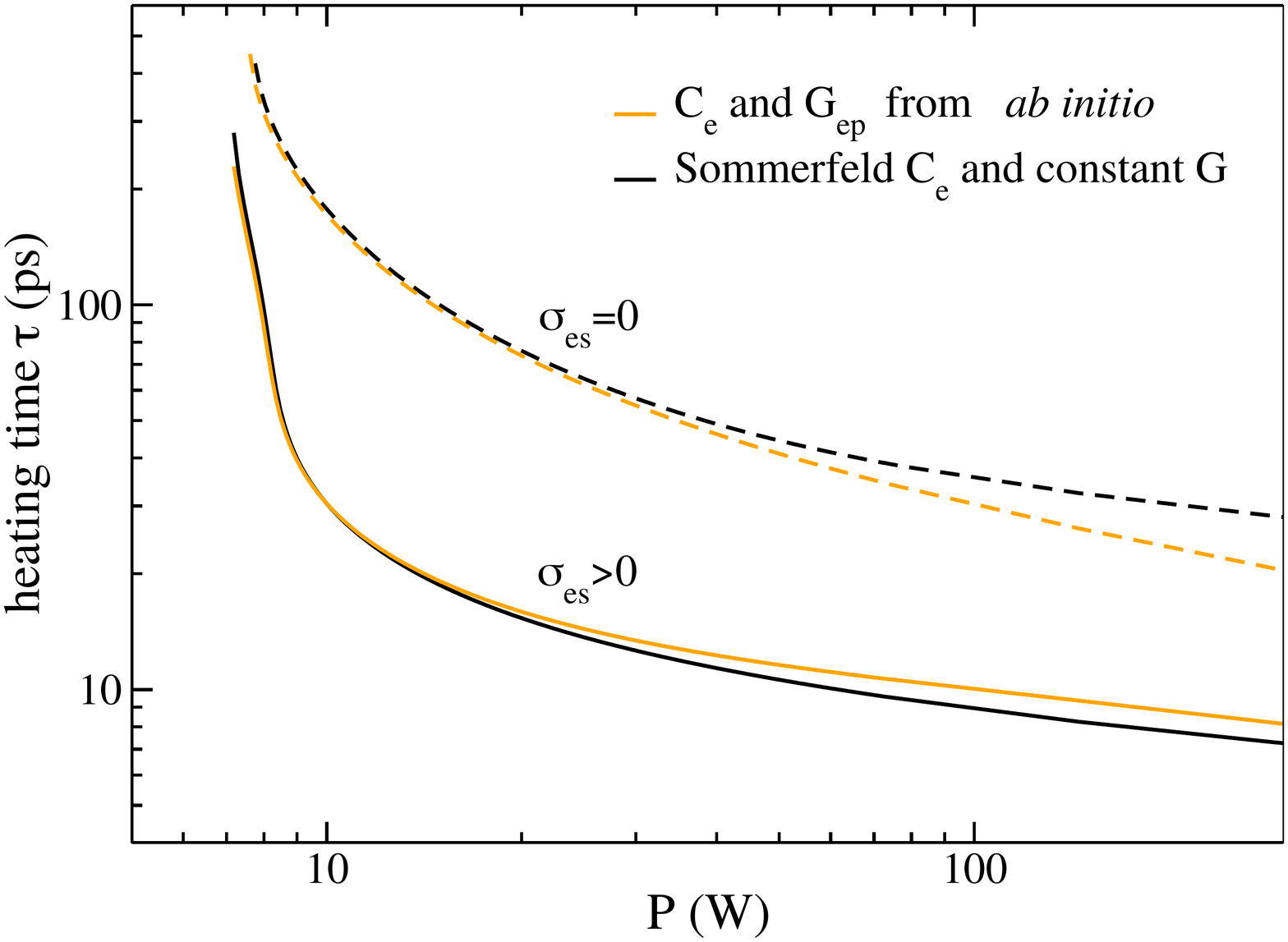}
\caption{\modif{Heating time to reach $\delT=25\,$K in two cases: 
constant $G$ and Sommerfeld heat capacity $\ce$~(black), 
vs {\em ab initio} values ~(orange). 
Dashed curves correspond to the case $\siges=0$ 
while continuous curves are the results with nonzero $\siges$. 
The nanoparticle considered here is a CSNP with $\Rc=40\,$nm and~$d=5\,$nm irradiated by a femtosecond laser pulse. }}
\label{fig:HeatingTimeNonLinear}
\end{figure}

To assess the effect of such non-linearities on the CSNP heating kinetics, 
we show in Figure~\ref{fig:HeatingTimeNonLinear}  
the heating time as a function of the laser power, 
when we take into account or not the non-linearities of both $G$ and $\ce$. 
It is clear that below $100\,$W, the non-linearities do not play a significant role, 
and the linear model used above provides an acceptable description of the heating of the CSNP. 
Beyond $100\,$W, a departure from the linear model may be seen, 
but it should be noted that in the case $\siges \neq 0$,
the deviations are limited to $15$ percents. 
It should also be reminded that above $100\,$W, we used a value of $\siges$ extrapolated from its low $\Te$ behavior, 
as we have no information on the electron-phonon cross coupling at electronic temperatures $\Te>3000\,$K. 
Exploring the heating kinetics of CSNP in this regime would require 
the experimental determination of $\siges$ under conditions of strong electron-phonon non-equilibrium. 
  
\section{Effect of finite diffusion in the metal}
\label{ap:diffusion}
\begin{figure}[htbp]
\includegraphics[width=8cm]{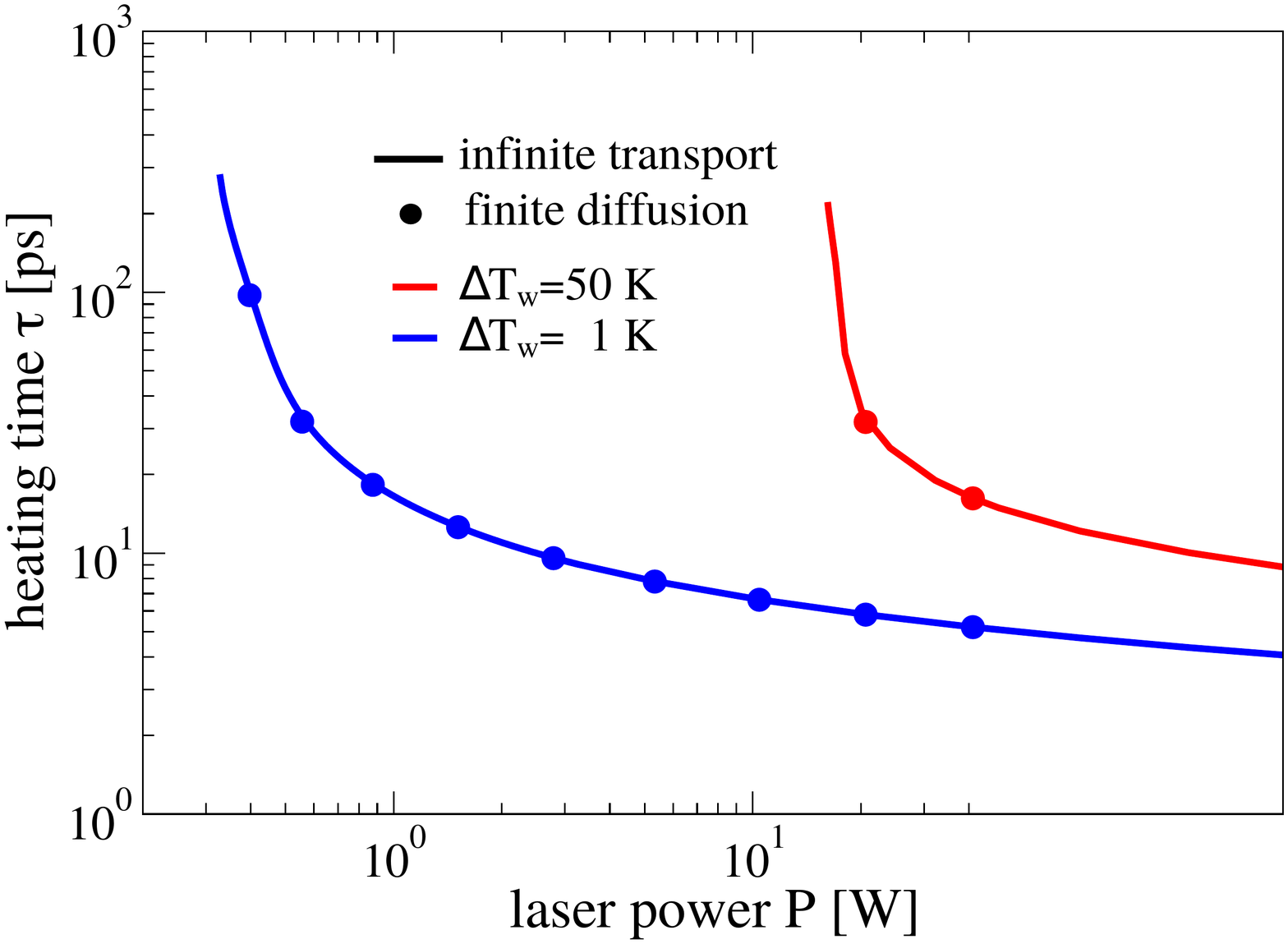}
\caption{Heating time to reach $\delT=1$ and $50\,$K 
assuming  for metal phonons instantaneous transport (lines) 
or finite diffusion (symbols). The nanoparticle considered here is a CSNP with $\Rc=40\,$nm and~$d=5\,$nm. }
\label{fig:HeatingTimeDiffusion}
\end{figure}

The model described by Eqs.~\eqref{eq:VarTe}-\eqref{eq:VarTw} assumes that 
the  diffusion of heat carriers in the metal is infinitely fast. 
The underlying reason is the high thermal conductivity of the gold free electrons (see Tab.~\ref{tab:thermophysical}), 
suggesting that the temperatures in the metal can be approximated as uniform. 
While this is a good approximation for electrons, 
it might not be  the case for phonons, 
since  their thermal diffusivity $\difp$ is significantly smaller.  
Their  diffusion  time over a typical length of $40\,$nm 
is $1250\,$ps, which is two orders of magnitude larger than that for electrons. 
We have checked how heat transfers are affected 
when the diffusion of  phonons in the metal is finite. 
As demonstrated in Fig.~\ref{fig:HeatingTimeDiffusion},  
the heating times are exactly the same with an infinite transport or finite  diffusion of phonon inside the metal. 
This indicates that our approximation is legitimate. 

\begin{acknowledgments}
We acknowledge illuminating discussions with P.~Maioli, A.~Crut, F.~Amblard and \modif{T. Niehaus}. 
JL thanks funding from DGAPA UNAM. 
SM acknowledges financial support from the FET Open project EFINED under the grant agreement 766853.
\end{acknowledgments}


\end{document}